\documentclass[showkeys]{revtex4}
\usepackage{epsfig}

\def\lsim{\mathrel{\mathop
  {\hbox{\lower0.5ex\hbox{$\sim$}\kern-0.8em\lower-0.7ex\hbox{$<$}}}}}
\def\gsim{\mathrel{\mathop
  {\hbox{\lower0.5ex\hbox{$\sim$}\kern-0.8em\lower-0.7ex\hbox{$>$}}}}}
\newcommand{\form}[1]{(\ref{#1})}
\newcommand{\be}{\begin{equation}}
\newcommand{\ee}{\end{equation}}
\newcommand{\ba}{\begin{array}}
\newcommand{\ea}{\end{array}}
\newcommand{\bea}{\begin{eqnarray}}
\newcommand{\eea}{\end{eqnarray}}
\newcommand{\half}{{1\over2}}

\newcommand{\x}{{\bf x}}
\newcommand{\X}{{\bf X}}
\newcommand{\z}{{\bf z}}
\renewcommand{\k}{{\bf k}}
\newcommand{\K}{{\bf K}}
\renewcommand{\t}{\tau}
\newcommand{\tnl}{\tau_{\rm nl}}
\newcommand{\tsb}{\tau_{\rm sb}}
\newcommand{\kd}{\sqrt{k\over2}}
\newcommand{\dk}{{1\over\sqrt{2k}}}
\renewcommand{\Re}{{\rm Re}\,}
\renewcommand{\Im}{{\rm Im}\,}
\newcommand{\ai}{{\rm Ai}}
\newcommand{\bi}{{\rm Bi}}
\newcommand{\cc}{^\dagger}

\newcommand{\uni}{1\!{\rm l}}
\newcommand{\Hint}{ {\mathbf H}_{\mbox{\small int}}} 
\newcommand{\fii}{y}
\newcommand{\ti}{\t_i}
\newbox\pippobox

\begin{document}

\preprint{CERN-TH/2002-156\\ IFT-UAM/CSIC-02-30\\ \hepph{0208228}}

\title{Symmetry Breaking and False Vacuum Decay after Hybrid Inflation}

\author{ Juan Garc\'{\i}a-Bellido} 
\affiliation{Departamento de F\'\i sica Te\'orica \ C-XI and Instituto de F\'\i sica Te\'orica \ C-XVI, 
\mbox{Universidad Aut\'onoma de Madrid, Cantoblanco, 28049 Madrid, Spain}\\ 
 }
\affiliation{TH-Division CERN, CH-1211 Geneve 23, Switzerland}
\author{Margarita Garc\'\i a P\'erez}
\affiliation{TH-Division CERN, CH-1211 Geneve 23, Switzerland}
\author{Antonio Gonz\'alez-Arroyo}
\affiliation{Departamento de F\'\i sica Te\'orica \ C-XI and Instituto de F\'\i sica Te\'orica \ C-XVI, Universidad
Aut\'onoma de Madrid, Cantoblanco, 28049 Madrid, Spain}

\begin{abstract} We discuss the onset of symmetry breaking from the false
  vacuum in generic scenarios in which the mass squared of the
  symmetry breaking (Higgs) field depends linearly with time, as it
  occurs, via the evolution of the inflaton, in models of hybrid
  inflation.  We show that the Higgs fluctuations evolve from quantum
  to classical during the initial stages.  This justifies the
  subsequent use of real-time lattice simulations to describe the
  fully non-perturbative and non-linear process of symmetry breaking.
  The early distribution of the Higgs field is that of a smooth
  classical gaussian random field, and consists of lumps whose shape
  and distribution is well understood analytically. The lumps grow
  with time and develop into ``bubbles'' which eventually collide
  among themselves, thus populating the high momentum modes, in their
  way towards thermalization at the true vacuum. With the help of some
  approximations we are able to provide a quasi-analytic understanding
  of this process.
\end{abstract}


\maketitle

\section{Introduction}

The problem of symmetry breaking in quantum field theory (QFT) has been  with
us for several decades. In the context of cosmology, it has
usually been associated with thermal phase transitions~\cite{KL}, and
the production of topological defects~\cite{KZ}. Understanding the way
in which the order parameter associated with the breaking of the
symmetry evolves from a symmetric state (the false vacuum) to a broken
state (the true vacuum) is non-trivial~\cite{CHKMPA,BdV,BHP,AB,Smit}. Only
recently has this problem been addressed in the context of symmetry
breaking at zero temperature, at the end of a period of hybrid
inflation~\cite{hybrid,GBL}, in the so-called process of tachyonic
preheating, i.e. spinodal instability, in the context of preheating
after inflation~\cite{KLS}. There, {\em classical} evolution equations
have been solved with real-time lattice simulations, developed for
studying the problem of preheating~\cite{TK,PR,FK}, which include all the
non-perturbative and non-linear character of the phase transition. It
was found that symmetry breaking occurs typically in just one
oscillation around the true vacuum~\cite{GBKLT}, most of the false
vacuum energy going into gradient modes, rather than kinetic energy.

However, the problem of the transition from a false {\em quantum} vacuum
state at zero temperature (as occurs at the end of a period of
inflation) to the true {\em quantum} vacuum state full of radiation at a
certain temperature has not been fully addressed yet. Most of the
previous approaches refer to the decay from a false vacuum state at
finite temperature, relying on the Hartree or large N approximation
\cite{CHKMPA,BdV,AB,Smit}. Only recently, the zero temperature problem
was addressed within the classical approximation but mainly for the case
of an instantaneous quench~\cite{GBKLT}, which can lead to
cosmologically interesting particle production~\cite{GBRM}. Tachyonic
preheating was recently studied beyond the quench approximation in
Ref.~\cite{CPR}.

We will argue that symmetry breaking proceeds through a state in which
the relevant degrees of freedom are semi-classical infrared modes,
which can be described in a non-perturbative and non-linear way with a
classical effective field theory, whose classical equations of motion
can be solved numerically in the lattice and thus allow us to study
the fully non-perturbative out of equilibrium process of symmetry
breaking.

The quantum to classical transitions of field Fourier modes have been
addressed before in the context of inflation \cite{GuthPi,PS,KPS,LPS},
where it is mandatory to understand the transition from quantum
fluctuations of the inflaton field during inflation to the classical
metric fluctuations on super-horizon scales, since they are believed
to be responsible for the observed temperature anisotropies in the
cosmic microwave background, as well as the scalar density
perturbations giving rise to galaxies and large scale structure
formation. The use of the classical approximation to study the process of 
preheating after inflation has been proposed in \cite{TK}, and, in a context 
similar to ours, it has been recently used in \cite{GBKLT,CPR,Smit}.

In this paper we will use such a well developed formalism to study the
first instances of a generic symmetry breaking process, i.e. the
conversion of quantum modes of the symmetry breaking field (the QFT
order parameter, generically called the Higgs field) into a classical
gaussian random field whose subsequent non-linear evolution equations
can be solved with lattice simulations. We give here a self-contained 
presentation of the conditions under which classical behaviour holds,
and apply it to the analysis of the false vacuum decay after inflation, 
with specific initial conditions. 

Our final aim is to study the process of electroweak symmetry
breaking, and the possibility of realizing baryogenesis at the
electroweak scale, via the non-equilibrium process of preheating
after inflation~\cite{GBKS,KT,RSC}. Therefore, this paper is
intended as the first one in a series, in which we will progressively
incorporate more complexity, i.e.  gauge fields, Chern-Simons,
CP-violation, etc., into the picture.  Of course, our results are
readily generalisable to any other phase transition that may have
occurred in the early universe at the end of a period of hybrid
inflation, e.g. at GUT scales.

The paper is organised as follows. In Section II we describe the initial
conditions for spontaneous symmetry breaking coming from a hybrid model
of inflation. The inflaton acts here like a background field whose
coupling gives a time-dependent mass to the Higgs. In Section III we
study the quantum evolution of the Higgs field in the linear
approximation from the bifurcation point. The Fourier modes decouple in
this approximation and can be studied as a quantum mechanical ensemble
of harmonic oscillators, both in the Heisenberg and the Schr\"odinger
picture. We then study, with the use of the Wigner function, the quantum
to classical transition of the Higgs modes. We show that each quantum
mode can be described exactly like a classical gaussian random field,
and give a prescription for computing the Weyl-ordered quantum
expectation values of operators in terms of classical averages over a
gaussian random field with the Wigner function as probability
distribution. We then define and characterise when a mode can be called
``quasi-classical''. 

In Section IV we give the exact solutions to the field evolution
equations of the Higgs in the linear approximation in terms of Airy
functions, and show that soon after the bifurcation the infrared modes
become quasi-classical according to the definition of the previous
section. This analysis follows closely, although in greater detail, what
has been studied previously in the literature~\cite{ABC,JGB,CPR}. In
Section V we analyse the inclusion of the non-linear terms in the
quantum evolution within perturbation theory.  We give a prescription
for treating the ultraviolet divergences and to renormalise the
parameters of the theory. This leads to a regular probability
distribution to be used for a classical field description, which matches
the renormalised quantum expectation values of the Weyl ordered
products.  The matching is done at a time in which the infrared modes
have grown sufficiently to be well described as classical modes.  This
occurs well before non-linearities are important, and therefore our
gaussian approximation is valid, in a similar spirit as that of
Ref.~\cite{Smit}. The quantum ultraviolet modes, on the contrary, can
be thought as integrated out, and used to renormalise the parameters of
the classical theory.  We might then interpret our classical field
distribution as an effective theory for the long wave-length modes. We
also estimate the time at which symmetry breaking sets in.
 
In Section VI we describe the methodology to be used to take care of the
full non-linear evolution of the system. The initial space-time
structure of the classical Higgs field is analysed in VI.A. Being a
gaussian random field, it can be described in a similar way to the
matter density field whose fluctuations give rise to galaxies and large
scale structure via gravitational collapse~\cite{BBKS}.  The Higgs field
is found to possess, at symmetry breaking, an inhomogeneous spatial
distribution made of lumps, whose shape and initial evolution can be
well understood analytically.  The space-time inhomogeneous character of
symmetry breaking in the Higgs-inflaton system has also been reported in
Ref.~\cite{CPR} for a one component Higgs model. For the complete
non-linear dynamics of the full Higgs-inflaton system we make use of
lattice real-time evolution methods. The details of our procedure, its
connection and difference with lattice methods used by other authors,
and a detailed check of the validity of the approximations used, are
described in Section VI.B.

Finally, in Section VII we present the results of this stage of the
evolution of the system. The lumps mentioned in the previous sections
grow and, once its center reaches the Higgs vacuum expectation value,
invaginate and create an approximately spherically symmetric
``bubble'' which expands at a very high speed. Meanwhile, the center
of the lump-bubble continues to oscillate with decreasing amplitude,
leading to secondary bubbles. All these phenomena can be well
understood with the help of some approximations which reduce the full
non-linear equations to a one or two dimensional partial differential
equation of a single scalar field. This simplified picture matches
qualitatively and (to a high degree) quantitatively the results of the
lattice simulations. Eventually, bubbles centred at different points
collide and transfer most of their potential and kinetic energy to
gradient energy, thus populating the higher momentum modes. This
process leads to complete symmetry breaking and (classical)
thermalization. The whole history of the system is illustrated by
following the evolution of 2D sections of a particular configuration.
We also show histograms for the field values of both Higgs and
inflaton, which start in the false vacuum and are seen to end up
peaked around the true vacuum. In Section VIII we draw our conclusions
and describe the future directions in which this work can be extended,
first by including the production of SU(2) gauge fields and afterwards
by studying the rate of sphaleron transitions that may give rise to a
non-negligible amount of baryons.

We have added three appendices.  In Appendix A we describe the
formalism of squeezed states following \cite{PS,KPS}, which can be
applied to the initial stages of the Higgs evolution in the linear
regime, and gives rise to the semi-classical nature of the long
wavelength modes.  In Appendix B we compute the Wigner function for
the evolved gaussian initial vacuum state \cite{PS,KPS,LPS}, and show
explicitly the squeezing of the infrared modes. We give a definite
condition for characterising the moment in which a mode can be treated
as quasi-classical. In Appendix C we give the details of the
perturbative calculations of the non-linear evolution of our system,
both at a classical and quantum-mechanical level.

\section{The Higgs field at the end of hybrid inflation}

The precise model of hybrid inflation will not be important for our
purposes here. However, for concreteness we will implement it in the
context of a super-symmetric extension of the standard model in which
the radiative corrections are responsible for the running of the
inflaton field during the few ($\sim 5-10$) e-folds necessary to cool
the universe so that the electroweak symmetry breaking (EWSB) occurs
at zero temperature. The fluctuations responsible for CMB temperature
anisotropies and large scale structure come from a previous stage of
inflation, completely independent of this.  Moreover, since EWSB occurs at 
low energies, we can, and will in what follows, safely ignore the rate of 
expansion, $H\sim 10^{-5}$ eV, during symmetry breaking and treat the fields 
as if they were in Minkowski space.  In particular, one can
consider the super-symmetric hybrid model of Dvali, Shafi and
Schaefer~\cite{DSS}, where the super-potential fixes a relation
between the couplings, $g^2=2\lambda$. As we will see, this choice
simplifies some stages of the dynamics of symmetry breaking after
inflation~\cite{BGK}, but is not crucial. A study of the process of
tachyonic preheating after a variety of more general super-symmetric
models of inflation will be given in Ref.~\cite{GBKL}.

The hybrid model we are considering is a simple generalisation of the
Standard Model symmetry breaking sector, which consists of the Higgs
field, $\Phi=\half(\phi_0\,\uni+i\phi^a \t_a)$, with $ \t_a$ the Pauli
matrices, and an inflaton $\chi$, a singlet under SU(2). The inflaton
couples only to the Higgs, with coupling constant $g$. The scalar
potential has the usual Higgs term plus a coupling to a massive
inflaton,
\be\label{potential}
{\cal L} = (D_\mu\Phi)^\dagger\,D^\mu\Phi + \half(\partial_\mu\chi)^2 -
\lambda\Big(\Phi^\dagger\Phi - {v^2\over2}\Big)^2 - 
g^2\chi^2\Phi^\dagger\Phi - \half \mu^2 \chi^2 \,,
\ee
where $\,v=246$ GeV is the expectation value of the Higgs in the true
vacuum, $\mu$ is the mass of the inflaton in the false vacuum and
$m\equiv\sqrt\lambda\,v$. We are assuming implicitly that whenever
there is a contraction $O^\dag O$, we should take the trace over the
SU(2) matrices, i.e. $\Phi^\dag\Phi \equiv {\rm Tr}\, \Phi^\dag\Phi =
\half(\phi_0^2 + \phi^a\phi_a)\equiv |\phi|^2/2$. The Higgs mass in
the true vacuum is determined by its self-coupling: $m_{\rm H} \equiv
\sqrt{2\lambda}\,v$, while the mass of the inflaton in the true vacuum
is given by $m_{\rm I} \equiv gv\gg\mu$.

In this paper we will simplify the analysis of the dynamics by
omitting the $SU(2)$ gauge field and working with a generic Higgs
field with $N_c$ real components. We anticipate that the most
important conclusions of this paper are not affected by the
introduction of the gauge field, and leave for a forthcoming
publication the symmetry breaking dynamics in the presence of gauge
fields. For ease of notation we will drop the internal indices of the
Higgs field whenever all components behave in the same way. The
numerical simulations that will be presented correspond to a $N_c=4$
component Higgs field.

During hybrid inflation~\cite{hybrid} the Higgs field has a large
and positive effective mass squared due to its coupling to the inflaton field,
which slow-rolls down its potential valley. The  potential
for the coupled fields is the following
\be\label{pot}
V(\phi,\chi) = {\lambda\over4} \Big(|\phi|^2-v^2\Big)^2 + {g^2\over2}
|\phi|^2\chi^2 + \half\,\mu^2 \chi^2\,.
\ee
where the parameters in the potential depend on the number of Higgs
components $N_c$ as: $\lambda = \lambda_0/N_c$, $g^2 = g_0^2/N_c$ and
$v^2 = N_c \, v_0^2$, with this $m^2=\lambda v^2=\lambda_0 v_0^2$
is independent of the number of components.

It is the effective false vacuum energy $V_0 = \lambda v^4/4 \equiv
m^2v^2/4$ which drives the period of hybrid inflation. Inflation ends
when the inflaton homogeneous mode, $\chi \equiv \langle\chi\rangle$, 
slow-rolls below the bifurcation point $\chi = \chi_c \equiv 
m/g$, at which the Higgs is massless,
\be
m^2_\phi = m^2\,\Big({\chi^2\over\chi_c^2} - 1\Big)\,.
\ee
Below the critical point, the Higgs has a negative mass squared and
long wave modes will grow exponentially, driving the process of symmetry 
breaking~\cite{GBKLT}. The process by which the mass squared of the
Higgs goes from large and positive to large and negative is not 
instantaneous, but depends strongly on the velocity of the inflaton
at the bifurcation point,
\be
V \equiv {1\over m}\left|{\dot\chi\over\chi_c}\right|_{t_c}\,.
\ee
Typically the speed of the inflaton is such that the process takes
place in less than one Hubble time, a condition known as the
``waterfall'' condition~\cite{hybrid,GBL}, which ensures the absence of
a second period of inflation after the bifurcation point~\cite{GBLW}.
The actual value of $V$ depends very much on the model and the scale
of inflation, and we will treat it here as an arbitrary model parameter.
In this case, the effective mass of the Higgs across the bifurcation
point can be written as a time-dependent mass
\be\label{masst}
m^2_\phi(t) = - 2V\,m^3 (t-t_c) + {\cal O} (V^2(t-t_c)^2) \,.
\ee

Note that a similar situation arises in the case of simple extensions
of the standard model Higgs, in which radiative corrections (dominated
by the large top quark Yukawa coupling) induce the running of the
Higgs mass square from positive to negative thus providing a mechanism
for Electroweak symmetry breaking. The role of the running scale is
played here by the inflaton homogeneous mode. Alternatively, one can
envisage a secondary period of hybrid thermal inflation
\cite{LS,review} just above the electroweak scale, which lasted only a
few $e$-folds and super-cooled the false vacuum, leaving only the fast
rolling inflaton coupled to the Higgs. This short second period of
inflation would not affect the CMB anisotropies, but would provide a
natural initial condition for the growth of quantum fluctuation of the
Higgs field, as they evolve across the bifurcation point, toward
symmetry breaking.  

Let us consider the effective action for the Higgs field
$\Phi(\x,t)$ ignoring the self-coupling $\lambda$-term 
(we omit the internal indices of the Higgs field), 
\be
{\cal S} = \int d^3\x\,dt\,\half\left[(\dot\phi)^2 - (\nabla\phi)^2 
- m^2_\phi(t)\,\phi^2\right]\,,
\ee
where we have included the time-dependent mass \form{masst}, to linear
order, which is the only effect that the presence of the homogeneous
mode $\chi(t)$ of the inflaton field induces in the evolution of Higgs
quantum modes.  

We now define a new scale $M\equiv(2V)^{1/3}\,m$, and thus
redefine our coordinates as
\bea
\tau &=& M\,(t-t_c) \hspace{1cm} \longrightarrow \hspace{1cm}
\dot\phi = M\,\phi'\,,\\
\X &=& M\,\x \hspace{2cm} \longrightarrow \hspace{1cm}
\K = {\k\over M}\,.
\eea
where primes denote derivatives w.r.t. $\t$, and $\k$ is the wavenumber
associated with the Higgs Fourier modes, 
$$\Phi(\k,\t) = \int{d^3\x\over(2\pi)^{3/2}}\,\Phi(\x,\t)\,
\exp(-i\,\x\cdot\k)\,.$$
From now on, we will use $\x$ and $\k$ as the normalised position and
momentum coordinates, i.e. we will work in units of $M=1$.  We will
also denote the normalised Higgs quantum fluctuations 
by $y = \phi/M$, for which
the effective action is
\be
{\cal S} = \int d^3\x\,d\t\,\half\left[(y')^2 - (\nabla y)^2 +
\t\,y^2\right] \,.
\ee
We can define the conjugate momentum as
${\displaystyle p = {\partial{\cal L}\over\partial y'} = y'}\,,$
and thus the corresponding Hamiltonian becomes
\be
{\cal H} = \int d^3\x\,\half\left[p^2 + (\nabla y)^2 - \t\,y^2\right]\,.
\ee
In momentum space, the Hamiltonian becomes
\be\label{Hamiltonian}
{\cal H} = \int d^3\k\,\half\left[p(\k,\t)\,p^\dagger(\k,\t) +
(k^2-\t)\,y(\k,\t)\,y^\dagger(\k,\t)\right]\,.
\ee
The Euler-Lagrange equations for this field can be written in terms of
the momentum eigenmodes as a series of uncoupled oscillator equations:
\be
y''(\k,\t) + (k^2 - \t)\,y(\k,\t) = 0\,.
\ee

\section{Quantum evolution in the gaussian approximation}

In this section we will start the description of the quantum evolution
of the system assuming that we can neglect the non-linear terms which 
are proportional to $\lambda$. Our goal is to determine the precise conditions
under which the system evolves into a classical one. Our presentation will 
be general and applicable to any time dependent harmonic oscillator system
with time-dependent spring constant $\omega^2(k,\tau)$,
only in the next section we will apply this formalism to our particular 
problem ($\omega^2(k,\tau)= k^2-\t$). Our results overlap and coincide with 
Refs.~\cite{GuthPi,PS,KPS}.

\subsection{The Heisenberg picture}

In the Heisenberg picture the quantum system is described by means of the 
position $y(\k,\t)$ and momentum operators $p(\k,\t)$ corresponding to each 
oscillator. The canonical equal-time  commutation  relations for the fields
($\hbar=1$ here and throughout) in position and momentum space are
\be \label{commw}
\Big[y(\x,\t),\ p(\x',\t)\Big] = i\,\delta^3(\x - \x')\,,\hspace{1.5cm}
\Big[y(\k,\t),\ p(\k',\t)\Big] = i\,\delta^3(\k + \k')
\ee
Furthermore, hermiticity of the operators in position space 
imply the relations
$y^\dagger(\k,\t)= y(-\k,\t)$ and $p^\dagger(\k,\t)= p(-\k,\t)$

We will assume that at $\t=\t_0=0$, i.e. at the bifurcation point
$t=t_c$, the state of the system is given by the ground state of the
Hamiltonian with oscillator frequency $\omega(k,0)=k$. It is
then useful to express the position and momentum operators in terms of
creation-annihilation operators at that time:
\be\label{ypktw}\ba{l}{\displaystyle
y(\k,\t_0) = \dk\,\Big(a(\k,\t_0) + a^\dagger(-\k,\t_0)\Big)}\,,\\[3mm]
{\displaystyle
p(\k,\t_0) = -i\,\kd\,\Big(a(\k,\t_0) - a^\dagger(-\k,\t_0)\Big)}\,.
\ea\ee

The quantum operators satisfy the classical equations of motion, which
we will write down as a system of coupled first-order equations
\be
{d\over d\t} v(\k,\t) \equiv {d\over d\t} \pmatrix{p(\k,\t) \vspace{2mm} 
\cr y(\k,\t)}= \pmatrix{0 & -\omega^2(k,\t)\vspace{2mm}\cr
1 &0} \pmatrix{p(\k,\t) \vspace{2mm} \cr y(\k,\t)}
\ee
whose solution can be expressed as: 
\be
v(\k,\t) = {\mathbf M}(k,\,\t)\,v(\k,\t_0) \equiv
\pmatrix{ \sqrt{2\over k}\,g_{k1}(\t)&\sqrt{2k}\,g_{k2}(\t)\vspace{2mm}\cr  
-\sqrt{2\over k}\,f_{k2}(\t) & \sqrt{2k}\,f_{k1}(\t) }\,v(\k,\t_0)\,,
\ee
where $f_{k1}\equiv\Re f_k$ and $f_{k2}\equiv\Im f_k$,  with
$f_k(\t)$ a complex solution of the equation of motion,  
with initial conditions,
\be\label{fkw}
f_k'' + \Big(\omega(k,\t)\Big)\,f_k = 0\,, \hspace{1cm} f_k(\t_0) = \dk \,,
\ee
and
\be\label{gkw}
g_k \equiv g_{k1}+ig_{k2}=i\,f_k' \,, \hspace{1cm}  g_k(\t_0) = \kd \,. 
\ee
Note that 
since the motion is Hamiltonian (i.e. canonical), the determinant 
of ${\mathbf M}(k,\,\t)$ is
$\det\,{\mathbf M}(k,\,\t) = 1\,, \forall\t$, a condition that 
is equivalent to the Wronskian of Eq.~\form{fkw} being one
at all times, 
\be\label{wrk}
i\,(f_k'\,f_k^* - {f_k'}^* f_k) = g_k\,f_k^* + g_k^*\,f_k = 
2\,\Re(g_k\,f_k^*) = 1\,.
\ee

The previous formulae allow us to compute the expectation value of
products of fields at any time $\t$ in terms of the expectation values
of fields at time $\t_0$. Substituting \form{ypktw} into the
expression for the fields at time $\t$ we obtain
\be\label{ypkt0}\ba{l}
y(\k,\t) = f_k(\t)\,a(\k,\t_0) + f_k^*(\t)\,a^\dagger(-\k,\t_0)\,,\\[3mm]
p(\k,\t) = -i\,\Big(g_k(\t)\,a(\k,\t_0) - g_k^*(\t)\,
a^\dagger(-\k,\t_0)\Big)\,.
\ea\ee

The quantum information of the system is encoded in the expectation values 
of products of fields. For a gaussian field the only quantities needed to
describe the system are the two-point expectation values,
\be
\langle 0,\t_0| v_a(\k,\t)\, v_b(\k',\t') |0,\t_0\rangle = 
\Sigma_{a b}(k,\t,\t')\, \delta^3(\k+\k')\,,
\ee
where $|0,\t_0\rangle$ is the initial vacuum state satisfying 
$a(\k,\t_0)|0,\t_0\rangle=0\,,\forall\k$.
The value of this matrix at any pair of times can be expressed in terms 
of the matrix ${\mathbf M}$ and the corresponding expectation values at time 
$\t_0$ as follows:
\be
\Sigma(k,\t,\t')={\mathbf M}(k,\,\t)\,\Sigma(k,\t_0,\t_0)\,
{\mathbf M}^T(k,\,\t')\,.
\ee
The quantum initial condition on the state of the system at time $\t_0$ 
amounts to:
\be
\Sigma(k,\t_0,\t_0)= \pmatrix{ {k\over2} & - {i\over2}\vspace{2mm}\cr
{i\over2} & {1\over2k}}\,.
\ee
Note that this matrix is  hermitian, but neither real nor symmetric, and
its determinant vanishes. The imaginary part results from the equal time 
commutation relations and does not depend on the particular state of the 
system. The real  symmetric part alone characterises completely the
state. 

Let us conclude this section by giving the expression of
the equal time expectation values at any other time:
\be
\label{sigmat}
\Sigma(k,\t,\t)=\pmatrix{ |g_k(\t)|^2 &  F_k(\t) - {i\over2}\vspace{3mm}\cr
 F_k(\t) + {i\over2} &  |f_k(\t)|^2 }
=\pmatrix{ |g_k(\t)|^2 &  -i\,\Omega_k^*(\t) |f_k(\t)|^2\vspace{3mm}\cr
 i\,\Omega_k(\t) |f_k(\t)|^2 &  |f_k(\t)|^2 }
\ee
with
\bea\label{omegak}
\Omega_k(\t) &=& {g_k^*(\t)\over f_k^*(\t)} =
{1 - 2i\,F_k(\t)\over2|f_k(\t)|^2}\,, \\[2mm] \label{Fk}
F_k(\t) &=&
\Im(f_k^*\,g_k) \,.
\eea
As a consequence of the unit determinant of ${\mathbf M}(k,\,\t)$ one
concludes that the determinant of the symmetric (real) part of
$\Sigma(k,\t,\t)$ is time-independent and equal to $1/4$.
Note that using \form{omegak} we can rewrite the conjugate momentum as
\be \label{pyrel}
p(\k,\t) =  \bar p(\k,\t) + {F_k(\t) \over |f_k(\t)|^2} \, y(\k,\t)
\ee
with 
$$\bar p(\k,\t) = -{i\over |f_k(\t)|^2} \,\Big(f_k(\t)\,a(\k,\t_0) - 
f_k^*(\t)\,a^\dagger(-\k,\t_0)\Big),$$
a relation that will prove useful in the next section.

\subsection{The Schr\"odinger picture and the classical limit}

Let us go now from the Heisenberg to the Schr\"odinger representation,
and compute the initial state vacuum eigenfunction $\Psi_0(\t=\t_0)$.
We will follow here Refs.~\cite{GuthPi,PS,KPS}. 
In what follows we will denote operators in the Schr\"odinger representation
by $\hat y_\k \equiv y (\k,\t_0)$ and $\hat p_\k \equiv p (\k,\t_0)$.
The initial vacuum state $|0,\t_0\rangle$ is defined through the condition
\bea
\forall \k\,, \hspace{1cm}  \hat a(\k,\t_0)|0,\t_0\rangle = 
\left[\kd\,\hat y_\k + i\,\dk\,\hat p_\k \right]
|0,\t_0\rangle = 0 \,, \hspace{1cm} \nonumber\\
\left[y_k^0 + {1\over k}\,{\partial\over\partial{y_k^0}^*}
\right]\Psi_0\Big(y_k^0, {y_k^0}^*, \t_0\Big) = 0 \hspace{5mm} 
\Rightarrow \hspace{5mm} \Psi_0\Big(y_k^0, {y_k^0}^*, \t_0\Big) = 
N_0\,e^{- k\,|y_k^0|^2}\,, 
\eea
where we have used the position representation, $\hat y_\k =
y_k^0\,, \ \hat p_\k = -i\,{\partial\over\partial{y_k^0}^*}$, and
$N_0$ gives the corresponding normalisation.

We will now study the time evolution of this initial wave function
using the unitary evolution operator ${\cal U} = {\cal U}(\t,\t_0)$, 
satisfying ${\cal U}' = -i\, {\cal H \, U} $. The state
evolves in the Schr\"odinger picture as $|0,\t\rangle =
{\cal U}|0,\t_0\rangle$. We can make use of the result of the previous 
section to determine this state.  By inverting \form{ypkt0}  we find
\be
\hat a(\k,\t_0) = g_k^*(\t)\,\hat y(\k,\t) + i\,f_k^*(\t)\,\hat p(\k,\t)\,,
\ee
which acting on the initial state becomes, $\forall \k\,, \forall \t$,
\bea
{\cal U}\left[\hat y(\k,\t) + i\,{f_k^*(\t)\over g_k^*(\t)}\,\hat p(\k,\t)
\right]{\cal U}^\dagger{\cal U}|0,\t_0\rangle = 
\left[\hat y_\k + i\,{f_k^*(\t)\over g_k^*(\t)}\,\hat p_\k
\right]\!|0,\t\rangle = 0 \,, \nonumber\\ \Rightarrow \hspace{1cm} 
\Psi_0\Big(y_\k^0, {y_\k^0}^*, \t\Big) \, = \,{e^{-i\alpha}
\over\sqrt\pi\, |f_k(\t)|}\,e^{- \Omega_k(\t)\,|y_k^0|^2}\,, \label{wftw} 
\hspace{3cm}
\eea
with $\Omega_k(\t)$ given by Eq.~\form{omegak}.  
We see that the unitary evolution preserves the gaussian form of the
wave functional. The wave function \form{wftw} is called a 2-mode
squeezed state. The normalised probability distribution, for each mode $k$, 
\be\label{probw}
P_0\Big(y_k^0, {y_k^0}^*, \t\Big) = {1\over\pi\,|f_k(\t)|^2}\,
\exp\Big(- {|y_k^0|^2\over|f_k(\t)|^2}\Big)\,,
\ee
is a gaussian distribution, with dispersion given by $|f_k|^2$. This
agrees with the result obtained in the previous section in the
Heisenberg picture.  The phase $\alpha(k,\t)$ cannot be determined by
this method, but as we have seen it has no effect on the probability
distribution nor on the Wigner function, see below. However, from the
Schr\"odinger equation, $i\partial_\t\Psi_0(\t) = {\cal H}\Psi_0(\t)$, 
one can deduce that $\alpha'(k,\t)=(2|f_k(\t)|^2)^{-1}$.

We can also compute the occupation number, $n_k$,
\be\label{ocn1}
n_k(\t) = \langle 0,\t|  a^\dagger(\k,\t_0)  a(\k,\t_0)  |0,\t
\rangle =
{1\over2k}|g_k|^2 + {k\over2}|f_k|^2 - \half \,,
\ee
a quantity that is always positive definite.

We now address the problem of approximating the quantum evolution just
described by a classical evolution. For that purpose the vacuum
expectation values of products of position and momentum operators
should be recovered as ensemble averages of random
fields.  It is clear that for the non-interacting theory ($\lambda=0$)
that we are considering, such a classical random field should be
gaussian, with all the information encoded in the real expectation
values of products of two fields. Only the symmetrical part of
$\Sigma(k,\t,\t)$ is real, see \form{sigmat}, and thus a natural
candidate to be approximated by the classical gaussian random field.
Notice that this corresponds to matching Weyl-ordered (symmetrised in
$\hat y_\k$, $\hat p_\k\cc$) quantum expectation values of
operators through, in the Schr\"odinger picture,
\be
\langle 0,\t| G(\hat y_\k, \hat p_\k) 
|0,\t\rangle_W \equiv \langle G(y_\k,p_\k)\rangle_{\rm gs}
\ee
where $\langle 0,\t| G(\hat y_\k, \hat p_\k )|0,\t\rangle_W$ denotes
the quantum average of the Weyl-ordered operator in the state given by
the wave function \form{wftw}, and $\langle G(y_\k,p_\k)\rangle_{\rm
  gs}$ denotes the classical gaussian average. The latter is obtained
as an average over a gaussian ensemble with $y_\k$ and ${\displaystyle
  \bar p_\k \Big(\equiv p_\k - {F_k(\t) \over |f_k(\t)|^2} \,
  y_\k\Big)}$ independent gaussian variables with probability
distribution given by the Wigner function in phase space, see
\cite{PS,KPS}, equation \form{pyrel} and the Appendix B
\be
\label{wigner2}
W_{0k}(y,p) = {1\over\pi^2}\,\exp\left(
-{|y|^2\over|f_k|^2} - 4|f_k|^2\Big|p - {F_k\over|f_k|^2}\,y\Big|^2
\right)\ .
\ee
A very trivial illustration of this equality is given by the following 
symmetrised vacuum expectation value 
\be
\label{example1}
\half \, \langle 0,\t|\,  \hat y_\k\,\hat p_{\k'} + \hat p_\k \,\hat y_{\k'} 
\, |0,\t \rangle =  F_k(\t) \delta^3 (\k+\k')
\ee
while
\be
\langle \,y_\k\, p_{\k'} \,\rangle_{\rm gs} = \langle\,  y_\k 
\,(\,\bar p_{\k'} + {F_k(\t) \over |f_k|^2}\,
y_{\k'}\,)\, \rangle_{\rm gs} = {F_k(\t) \over |f_k|^2}\, 
\langle\, y_\k \,y_{\k'} \,\rangle_{\rm gs} = F_k(\t)\, \delta^3 (\k+\k')\,.
\ee 

Even though only the symmetrised expectation values are described by
the gaussian ensemble average, for the gaussian ground state of
\form{wftw} quantum expectation values with arbitrary ordering of
operators can also be computed. To be specific, the expectation value
of any operator $G(\hat y, \hat p)$, with any given ordering of $\hat
y$ and $\hat p$, can be rewritten as a lineal combination of
Weyl-ordered operators with coefficients proportional to the
commutator which is a time independent c-number; schematically
\be
\label{semexp}
\langle 0,\t| G(\hat y, \hat p ) |0,\t \rangle  = 
\langle G_0(y ,  p) \rangle_{\rm gs}  \,+ \sum_{n\ge 1} \,(i\hbar)^n\,  
\langle G_n(y ,  p)\rangle_{\rm gs} \,,
\ee
where we have introduced $\hbar$ as an expansion parameter 
to make explicit the connection with the semi-classical
approximation. For instance, for the example in \form{example1} we would
obtain
\be
\langle 0,\t| \, \hat y_\k\,\hat p_{\k'} \, |0,\t \rangle
= \Big(F_k(\t) - {i\over 2}\Big) \, \delta^3 (\k+\k')
\ee
In this spirit, a quasi-classical state can be defined as a state for
which the leading term in equation \form{semexp} dominates, and
quantum averages can be approximated by
\be
\label{quasic}
\langle 0,\t|\, G(\hat y , \hat p) \,|0,\t \rangle 
\approx \langle\, G_0(y,p)\,\rangle_{\rm gs}.
\ee
This generically happens when $\langle 0,\t|\, \hat  p \,\hat y \,
|0,\t \rangle_W \gg |\langle 0,\t|\, [\,\hat p,\hat y\,]\,|0,\t \rangle|$, 
i.e. when the so-called WKB phase, $F_k(\t)$ in \form{Fk}, 
verifies $|F_k(\t)|\gg1$. For such a quasi-classical state the ambiguity 
in the ordering of operators is quantitatively negligible and 
classicality in the sense of \form{quasic} holds. As an illustration 
let us compute the following expectation value (for ease of notation
we have omitted the $\k$ dependence of the operators and the delta
functions)
\be
{1\over 2}\, \langle 0,\t|\,  \hat p^2 \, \hat  y^2 \, + \, \hat  y^2\,  
\hat p^2 \, |0,\t \rangle = 3 F_k^2(\t) -{1\over 4}.
\ee
In the classical approximation we would obtain
\be
\langle \, p^2  \,y^2  \,\rangle_{\rm gs} =  \langle  \,p^2  
\,\rangle_{\rm gs}  \,\langle  \,y^2  \,\rangle_{\rm gs} +   \, 
2 \langle \, p \, y  \,\rangle_{\rm gs}^2 =
|f_k(\t)|^2  |g_k(\t)|^2 +  2 F_k^2(\t) = 3 F_k^2(\t) + {1\over 4}
\ee
which reproduces the Weyl ordered part of the quantum result and is a very
good approximation as long as $|F_k(\t)|\gg1$.

This approach works as long as the theory is non-interacting and the
gaussianity of the quantum state is preserved by the evolution. 
In the problem at hand, we can assume this to be the case in the first 
stages of evolution before non linearities have set in, but not when 
the self-coupling term starts to
be relevant just before symmetry breaking.  However, as long as
$|F_k(\t)|\gg1$, the state can still be approximately described,
through the non-linear stages, via a classical  random field.
We will describe below in detail how this classicality follows for our
specific problem (see also the Appendices A and B for the precise
formulation of the squeezed states and the Wigner function formalism,
following Refs.~\cite{PS,KPS,LPS}).

\section{Exact solutions of the field evolution equations }

Let us apply now the above formalism to the case of the quantum
fluctuation modes of the Higgs at symmetry breaking. 
This case was studied previously in \cite{BM,ABC,JGB,CPR}.
For $\omega^2(k,\t)=k^2-\t$ the linear equation \form{fkw} for the quantum modes
of the Higgs field becomes
\be
f_k'' + (k^2-\t)\,f_k = 0\,, \hspace{1cm} {\rm with} \hspace{5mm}
f_k(\t_0=0) = \dk\,.
\ee
Its solution can be given in terms of Airy functions~\cite{AS}: 
\bea
f_k(\t) &=& C_1(k)\,\bi(\t-k^2) + C_2(k)\,\ai(\t-k^2) \,,\label{eqfk}\\[3mm]
g_k(\t) &=& i\,C_1(k)\,\bi'(\t-k^2) + i\,C_2(k)\,\ai'(\t-k^2) \,,
\label{eqgk}\\[2mm]
C_1(k) &=& -{\pi\over\sqrt{2k}}\,\Big[\ai'(-k^2) + ik\,\ai(-k^2)\Big]\,,\\
C_2(k) &=& {\pi\over\sqrt{2k}}\,\Big[\bi'(-k^2) + ik\,\bi(-k^2)\Big]\,,
\eea
which satisfy the Wronskian condition, $g_k\,f_k^* + g_k^*\,f_k = 1$,
where we have used the corresponding Wronskian for the Airy functions,
\be\label{WronskianAiry}
\pi\,[\ai(z)\bi'(z) - \bi(z)\ai'(z)] = 1\,.
\ee

We can then compute the occupation number, $n_k$, \form{ocn1}
and the imaginary part of the WKB phase, $F_k$, see \form{Fk}.

\subsection{Long wavelength quasi-classical modes}

Still remains to be computed the time at which fluctuations become
classical in the sense of Eq.~\form{quasic}. As we will see, the
field fluctuation modes will become quasi-classical as their wavelength
becomes larger than the only physical scale in the problem, the
time-dependent Higgs mass, i.e. $\lambda = 2\pi/k \gg 2\pi/\sqrt\t$.
In order to show this, let us take the limit $k^2 \ll \t$ for the long
wavelength modes in the exact solutions \form{eqfk} and \form{eqgk},
\bea
f_k(\t) &=& C_1(k)\,\bi(\t) + C_2(k)\,\ai(\t) \ 
\simeq \ C_1(k)\,w(\t)\,,\\[2mm]
g_k(\t)&=& i C_1(k)\,\bi'(\t) + i C_2(k)\,\ai'(\t) \
\simeq \ i\,C_1(k)\,w'(\t) \,,
\eea
where the function $w(\t)=\bi(\t)+\sqrt3\,\ai(\t)$ is the one appearing 
in the Appendix A, and we have used the fact that 
$C_2(k)=\sqrt3\,C_1(k)$ in the limit $k\to0$.
Using the large $z\gsim1$ approximation of the Airy functions, see
Ref.~\cite{AS},
\bea
\bi(z) \sim {1\over\sqrt\pi}\,z^{-1/4}\,e^{+{2\over3}z^{3/2}}\,,
&\hspace{1cm}
\ai(z) \sim {1\over2\sqrt\pi}\,z^{-1/4}\,e^{-{2\over3}z^{3/2}}\,,\\
\bi'(z) \sim z^{1/2}\,\bi(z)\,, \hspace{1.2cm}
&\hspace{2mm}
\ai'(z) \sim - z^{1/2}\,\ai(z)\,,
\eea
we conclude that the first terms in both $f_k$ and $g_k$ correspond to
the growing modes, while the second terms are the decaying modes, and
can be ignored soon after the bifurcation point.

We are now prepared to answer the question of classicality of
the modes. The wave function phase shift is given by
\be
F_k = \Im(f_k^*\,g_k)  \simeq
|C_1(k)|^2\, w'(\t) \, w(\t) \simeq {w'(\t)\over w(\t)} \, |f_k|^2 \simeq
|C_1(k)|^2\,{1\over\pi}\,e^{{4\over3}\t^{3/2}} \,,
\ee
which grows faster than exponentially at large time. 
On the other hand, the occupation number \form{ocn1} is
\be
2k\Big(n_k+\half\Big) = |g_k|^2 + k^2\,|f_k|^2 \simeq
\Big ( {w'(\t) \over w(\t)} \Big )^2 \,|f_k|^2
\simeq \sqrt\t\,|C_1(k)|^2\,{1\over\pi}
\,e^{{4\over3}\t^{3/2}}\,.
\ee
Therefore, we have
\be\label{classical}
|F_k(\t)| \simeq 
{2k\,n_k(\t)\over(w'/w)} \gg 1 \hspace{5mm}
\longrightarrow \hspace{5mm} E_k(\t) \simeq k\,n_k(\t) \gg |m(\t)|\,,
\ee
that is, $|F_k|\gg1$ whenever the energy of the mode $E_k$ is much
greater than the Higgs mass, computed as the instantaneous curvature
of the Higgs potential.

Notice in particular that under the condition $|F_k|\gg1$, the 
momentum and field eigenmodes are related by
$$ g_k(\t) = \Omega^*_k(\t) f_k(\t) \ \longrightarrow \
i {F_k(\t)\over |f_k(\t)|^2 } f_k(\t)  \,.$$
In terms of the gaussian random fields, the momentum distribution of
the Wigner function becomes a delta function, ${\displaystyle
\delta\Big(p-{F_k(\t)\over |f_k|^2} y\Big)}$, see Appendix B.

\begin{figure}
\includegraphics[width=10cm,angle=-90]{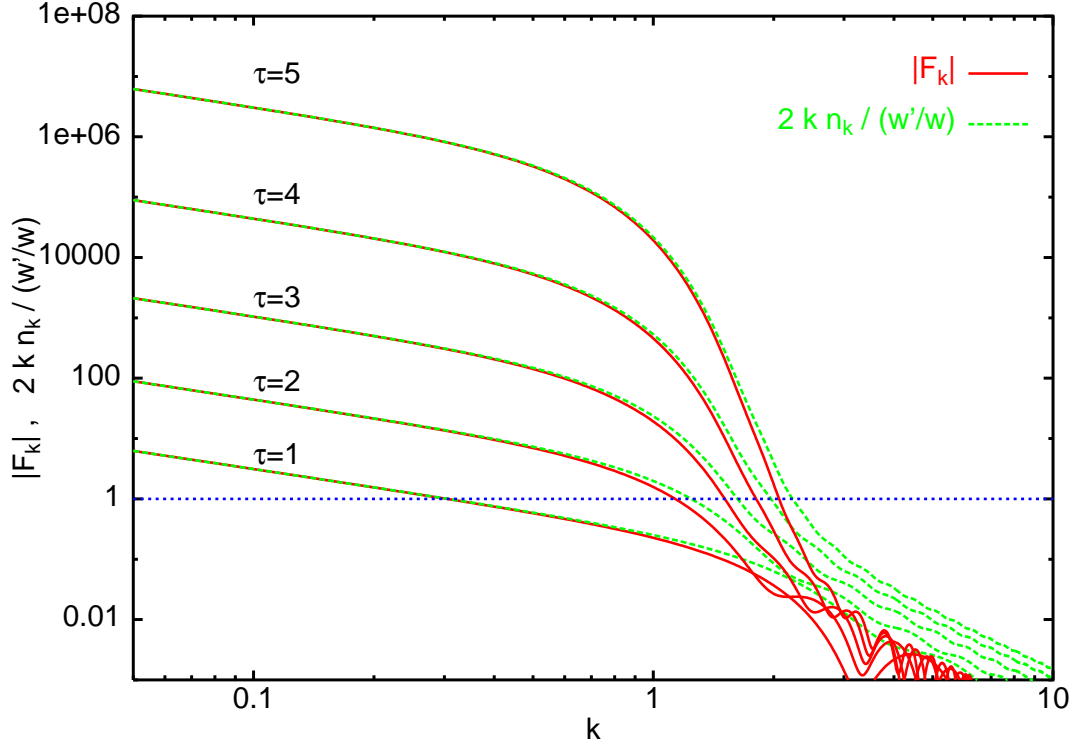}
\caption{We compare
the phase $|F_k|$ with the occupation number for different times, in
the whole range of interest in momenta $k$. Clearly, for large
times $\t\gg1$, the two coincide, as discussed in the text. Note that,
after $\t \simeq 2$, all long wavelength modes are essentially
classical, $|F_k|\gg1$.}
\label{Fknk}
\end{figure}

\begin{figure}
\includegraphics[width=10cm,angle=-90]{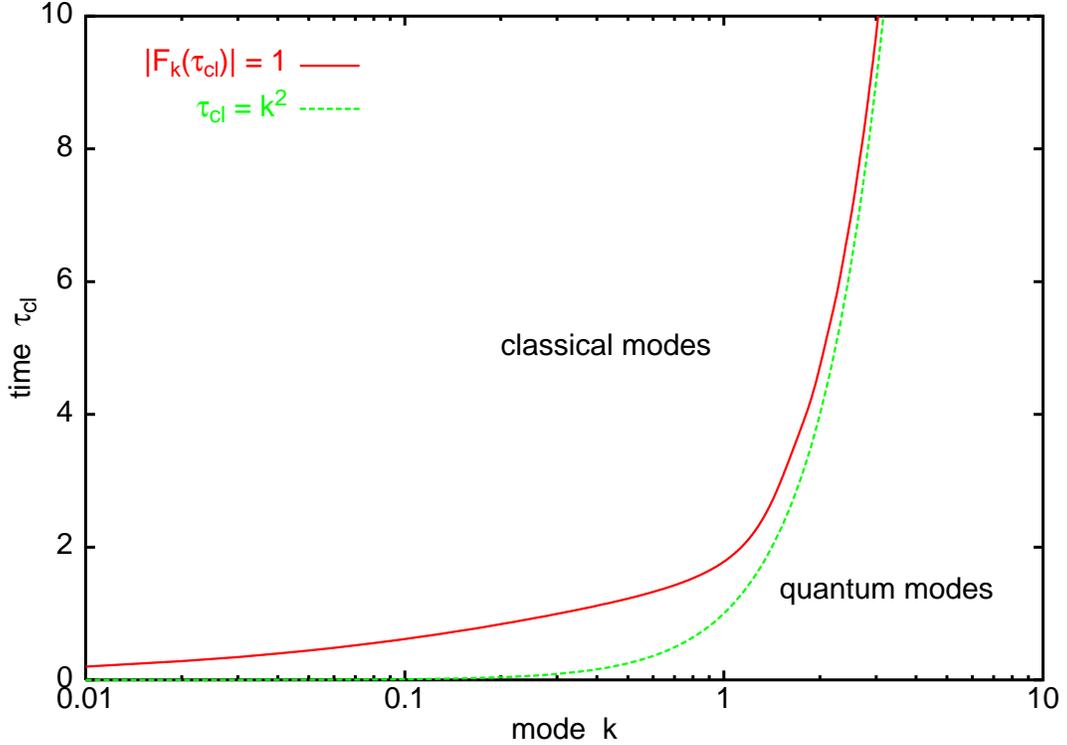}
\caption{The time
for which a given mode $k$ can be treated as classical 
$|F_k(\t_{\rm cl})| \equiv 1$ is above the line in this figure.
It is clear that long wavelength modes with $0 < k \lsim 1$ become
classical very early, at $\t_{\rm cl} \simeq 2$, while there
remains, at any given time, a high energy spectrum of quantum modes,
for $k \gg 1$.}
\label{Fk1}
\end{figure}

We show in Fig.~\ref{Fknk} the exponential growth of the phase
$F_k(\t)$ as a function of momenta $k$, for different times. These
plots were obtained using the exact Airy function solutions. Note that
in the limit of large wavelengths $k^2 \ll \t$, it is indeed verified that
$|F_k(\t)| \simeq 2k\,n_k(\t)(w/w') \gg 1$, as stated above.

We can now compute the time for which a single mode $k$ becomes
quasi-classical, in the sense \form{classical}. We have confirmed that
after $\t\simeq 2$ modes with $0 \leq k < \sqrt{\t} $ , which, as we
will see later, is the range of interest, have become
quasi-classical. We have drawn the line separating classical from
quantum modes in Fig.~\ref{Fk1}, as a function of the mode $k$. The
high energy part of the spectrum always remains in the quantum vacuum,
as expected. For $\t \geq 2$ the line separating classical and quantum
modes is approximately described by $k=\sqrt{\t}$.

\section{Non-linear quantum evolution and symmetry breaking}\label{nonlin}

To address the issue of symmetry breaking after inflation it is
essential to incorporate the non-linear effects proportional to
$\lambda$. A full non-perturbative quantum treatment is beyond reach.
However, we have seen in the previous section that the dynamics in the
absence of non-linear terms gives rise to a fast growth of the
amplitude of the low-lying momenta, leading to wave-functions which
are squeezed (quasi-classical in our language). We argue that even when
the interaction is switched on the dynamics of these modes dominates
the evolution of the system (at least during the first stages), and
that this dynamics is described by classical field theory. The
argument does not apply to higher momentum modes which sit largely in
the quantum mechanical ground state.  However, in quantum field
theory, high momentum modes, although small, do not give negligible
contribution to observables.  Actually, naively their contribution is
divergent.  Nevertheless, we argue that the main contribution of the
small quantum mechanical high-momentum modes sits in the
renormalization of the constants to be used in the classical theory.

It is possible to partially test this scheme in perturbation theory.
Already at this stage the problem of infinities and renormalization
arises~\cite{BHP}. In this section we will summarily analyse this
issue, relegating the details of the calculations to the Appendix
C. As we will see, for the program to be consistent one has to allow
for a renormalization of the speed $V$ of the inflaton at the
bifurcation.

In the standard setting, infinities in observables occur through the
contribution of the infinite tower of momentum states.  Introducing a
cut-off in the problem makes the results finite, but cut-off
dependent. It turns out, however, that in renormalizable theories, the
only surviving effects of the cut-off at scales much smaller than
itself are the modification of the constants of the theory. This
allows the process of renormalization in which we recover uniqueness
of the theory at the expense of taking this constants from experiment.
We will now reexamine this problem for our time-dependent situation.
Several research groups have investigated this problem in the past in
different contexts, see \cite{CHKMPA,BdV,BHP,ASm}.

All the physical content of the theory is contained in the expectation
values of products of the field operator at equal or different
space-time points (we use the Heisenberg picture and expectation values
should be understood as taken in the vacuum at $\t=\t_0$):
\begin{equation}
\langle y(\tau_1,\x_1) \ldots y(\tau_n,\x_n)\rangle\,.
\end{equation}
By differentiating with respect to $\tau$ one can obtain expectation
values of products of $\fii$ and $p$.  If we were to compute these
quantities in the gaussian (non-interacting) theory, we would obtain,
via Wick's theorem, a sum over all pairings of a product of factors
associated to each pair 
\begin{equation}
G^{(0)}(\tau,\tau',\x-\x')\equiv
\langle y_0(\tau,\x) y_0(\tau',\x')\rangle=\int
\frac{d^3\k}{(2 \pi)^3} e^{i\k(\x-\x')}\, 
f_k(\tau)f^*_k(\tau')\,,
\end{equation}
where $y_0$ denotes the gaussian field for $\lambda=0$. The gaussian 
two-point function is the Fourier transform of $\Sigma_{2 2}(k,\t,\t')$ 
and is finite provided $\x \ne \x'$ and/or $\tau \ne \tau'$. 

The correlation functions at different times (i.e. Wightman functions)
can be computed in perturbation theory by the method described in the
Appendix C. Wightman functions are complex and unlike Feynman
Green-functions (time-ordered products) depend on the order of the
operators. According with our criterion for the gaussian case, we will
consider Weyl-ordered (symmetrised) products to make the matching with
the classical theory.  If we now consider the symmetrised two-point
function
\begin{equation}\label{twopointA}
\langle y(\x,\t)  y(\x',\t')\rangle_W= \int
\frac{d^3\k}{(2 \pi)^3} e^{i\k(\x-\x')}\,\hat{G}(k,\t,\t')
\end{equation}
we can compute  it to first order in $\lambda$. The result is
\bea\label{twopointB} 
\hat{G}(k,\t,\t')=\Re[f_k(\t)f_k^*(\t')] &+&
\,2 (N_c+2)\lambda \int_0^\t ds\,A(s)\,\Im[f_k(\t)f_k^* (s)]\,
\Re[f_k(\t')f_k^* (s)]\\ \nonumber 
&+& \,2 (N_c+2)\lambda \int_0^{\t'} ds\,A(s)\,\Im[f_k(\t')f_k^* (s)]\,
\Re[f_k(\t)f_k^* (s)]\,.
\eea
where $N_c$ denotes the number of components of the Higgs field. 
The quantity $A(s)$ gives the contribution of the tadpole sub-diagram, 
i.e. the two-point function at equal times and zero distance, and is 
given by 
\be\label{phi2A}
A(\t)\equiv G^{(0)}(\tau,\tau,0)= \int {d^3\k\over(2\pi)^3}\,|f_k(\t)|^2 =
{1\over2\pi^2}\,\int{dk\over k}\,P(k,\t) \,,   
\ee
where the power spectrum is defined as $P(k,\t) = k^3|f_k(\t)|^2$.
This quantity is ultraviolet divergent. The structure of the divergence can
be deduced by analysing the large $k$ behaviour of the integrand.
Using our previous expressions (with $z=k^2-\t$) and the asymptotic 
behaviour of Airy functions~\cite{AS} we get
\bea
|f_k(\t)|^2 \simeq \,
{1\over2k}\left[ 1 + {\tau\over2k^2} \Big(1-{\sin(2k\t)\over 2k\t}\Big)+ 
{\cal O}(\t^2 )\right]  \,,\\ \label{pdiv} 
P(k,\t) = k^3|f_k(\t)|^2 \sim {k^2\over2} 
+ {\t\over4} \Big(1-{\sin(2k\t)\over 2k\t}\Big)+
{\cal O}\Big({\t^2\over k}\Big)\,.
\eea
Thus $A(\t)$ has a time-independent quadratic divergence and a linear 
in time logarithmic divergence. 

Before explaining how can one deal with the divergence, we comment
that Eqs.~(\ref{twopointA}-\ref{twopointB}) coincide precisely with the
calculation of the expectation values of the product of classical
random field to the same order in perturbation theory. Divergences are
hence present in both the quantum and the classical theory. Details
of this calculation are also shown in Appendix C.

We now address the problem of infinities that have occurred at this
level.  In the standard quantum theory the procedure is well-known.
The calculation can be done using some regulator to cut off the
contributions of high momenta, but this has to be accompanied by the
addition of counter-terms in the interaction Hamiltonian.  For the
theory to be renormalizable these counter-terms should have the same
expression as those appearing in the Hamiltonian (free or interacting)
but with coefficients which are cut-off dependent and proportional to
some power of $\lambda$. This addition should get rid of infinities.
Note that in our case a counter-term of the form
\be
- {N_c+2 \over 2} \lambda  \Big(\delta_1(\Lambda)  + \t\,
\delta_2(\Lambda)\Big) \, y_0^2(\x,\t) \,,
\ee
with $\delta_1$ and $\delta_2$ appropriately
chosen cut-off dependent functions, is able to subtract the
infinities encountered in $A(t)$. Regularising the integrals by introducing
a cut-off in momenta $k<\Lambda$ we then get:
\be
\label{areneq}
A_{\mbox{\tiny ren}}(\t,\mu)=A_{\mbox{\tiny reg}}(\t, \Lambda)
-\delta_1(\Lambda,\mu)-\t\,\delta_2(\Lambda,\mu)
\ee
To fix the arbitrariness introduced in the theory by the counter-term we 
must impose adequate renormalization conditions. As will be argued below,
one convenient possibility is to choose the counter-term as
\be
\label{match}
\delta_1 (\Lambda,\mu) + \t\,\delta_2(\Lambda,\mu) = 
{1 \over 2\pi^2} \int_\mu^{\Lambda} dk \, k^2 \, \Big( |f_k(\t=\mu^2)|^2 \, + 
(\t-\mu^2)\, 2\, F(\t=\mu^2)\Big) \, .
\ee
where $\mu$ denotes the characteristic mass scale of the problem
which for a given time $\t$ is precisely $\sqrt{\t}$. 
We will call this renormalization prescription, the fixed-time subtraction
scheme. Another possibility is a minimal subtraction scheme (not to be
confused with the MS scheme of dimensional regularization)
\be
A_{\mbox{\tiny ren}}(\t, \mu)=A_{\mbox{\tiny reg}}(\t, \Lambda)
-{1\over8\pi^2}\,\left(\Lambda^2 + \t\,\log{\Lambda\over\mu} 
\right)\,.
\ee
which differs from the previous one by finite terms
of the form $a +b \t$.
Actually, the renormalised quantity is obtained only after
taking the limit $\Lambda\rightarrow \infty$ in the subtracted
quantity,  but in practice taking $\Lambda$ sufficiently large is a
good approximation.

The fact that the structure of the counter-terms (or of the
divergence) has the same form as the terms already present in the
Hamiltonian, shows that our calculation is consistent at least to this
order. The infinities are re-absorbed in the values of the constants
of the theory. Here, in addition to the ordinary time-independent
subtraction we have a counter-term linear in $\t$, which can be
interpreted as a renormalization of $\chi-\chi_c$, i.e. of the
inflaton velocity $V$.  A different choice of scheme is compensated by
a finite renormalization of the parameters of our model.

Now we look back at the problem of approximating the result by a
classical random field. Since the regularised result to this order is
the same (for symmetric expectation values) a similar subtraction
procedure is necessary. There is certainly no problem to do so in
perturbation theory. However, in practice what we want to do is to be
able to match the renormalised quantum result by modifying the initial
spectrum of the classical field to be used as starting point for
the classical evolution. Notice that when $\t=\ti=\mu^2$ the   
value of $A_{\mbox{\tiny ren}}$ obtained with the fixed-time subtraction 
scheme, \form{match}, is exactly reproduced by truncating the initial spectrum 
at $\mu=\sqrt{\ti}$. This is a very natural choice from the
point of view of the classical approximation. As can be seen from
Figs. \ref{Fknk}, \ref{Fk1}, for large enough $\t$ the separation between 
quantum and classical modes sits indeed at $k \simeq \sqrt{\t}$.  
At a given time modes with momenta below $\sqrt{\t}$ have been
amplified while those above $\sqrt{\t}$ remain in the vacuum.
The amplification proceeds until some time $\tsb$, when $\langle \phi^2 
(\tsb)\rangle$ gets close to the vacuum expectation value $v^2$ and the 
field starts 
oscillating around the true vacuum. The dynamics of symmetry breaking is hence
expected to be governed by the low momentum modes with $k^2 < \tsb$
whose evolution can be described in the classical approximation
(as we will see below for a large range of parameters, $\tsb$ varies
only within the values $\tsb= 5 \pm 2$). The classical theory 
can then be seen, in a way analogous to what happens at high temperature 
\cite{ASm,Hight}, as an effective theory where momenta above $k_*=\sqrt{\tsb}$ 
have been integrated out. As far as modes above $k_*$ are not highly 
populated by rescattering and back-reaction this effective theory is expected 
to be valid and can be studied within the classical approximation.

In summary, our proposal is to fix  our classical field by matching its
correlation functions with the renormalised perturbative expression at
a time $\ti=\mu^2$ such that a sufficiently large number of momentum modes
have become classical but well before non linearities have set in.
The initial spectrum of the classical field will be cut-off at
$k=k_*=\sqrt{\ti}$. This eliminates the UV infinities of the classical theory.
If we  compare now with the calculation at one loop, we realise that
the parameters entering the classical theory are the renormalised parameters
in the fixed-time subtraction scheme \form{match}.
As we will see in what follows and in Section \ref{lattice} our results
are  fairly insensitive to the specific choice of $\ti$ within a
scaling window  below $\tsb$.

\begin{figure}
\includegraphics[width=10cm,angle=-90]{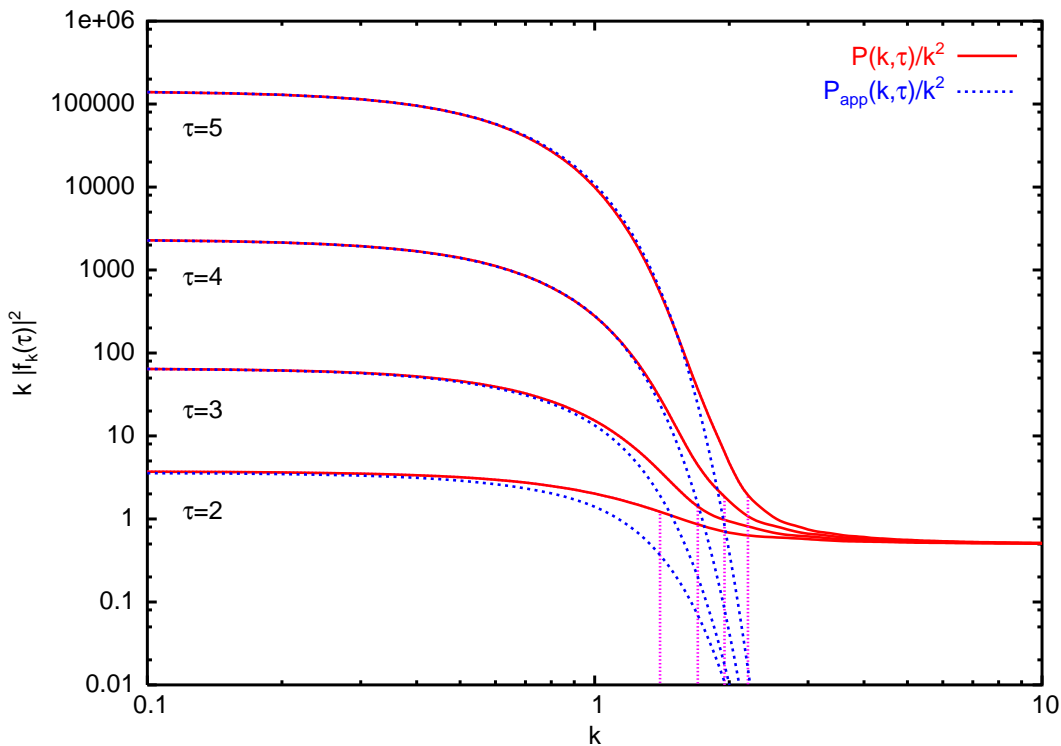}
\caption{The power
  spectrum of the Higgs quantum fluctuations, $P(k,\t)/k^2 \equiv k\,
  |f_k(\t)|^2$, at different times in the evolution. The dotted vertical
lines indicate the value of the cut-off, at $k=\sqrt{\t}$, where the classical 
spectrum is truncated. Also shown is the excellent approximation \form{Papp} 
in the region of long wave modes.}
\label{Pkreg}
\end{figure}

The validity of this approximation can be partially tested in
perturbation theory. A first check is the form of the power spectrum
\form{phi2A}. We have plotted in Fig.~\ref{Pkreg} the power spectra
$P(k,\t)$ divided by $k^2$, for four values of the normalised time $\t=2, 3, 4, 5$.
Clearly, the power spectra grow in time faster than an exponential, at a very
large rate in fact. We take as initial spectrum of the classical field at 
a given initial time $\ti$, the exact power spectrum cutoff 
at $k=\sqrt{\ti}$. As seen in the figure this encompasses almost all the 
physically relevant low momentum modes for  $\ti \gsim 2$. 
 
\begin{figure}
\includegraphics[width=12cm,angle=-90]{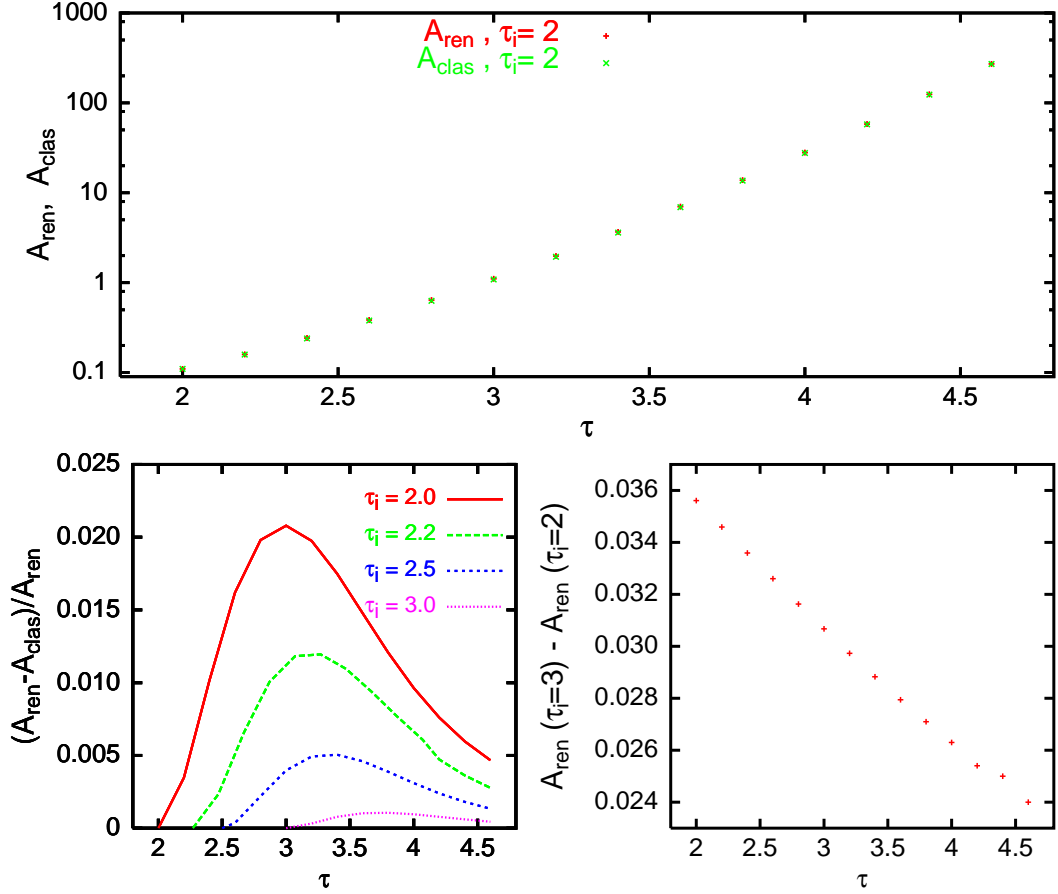}
\caption{Top:
  Comparison between $A_{\mbox{\tiny ren}}(\t,\mu\!=\!\sqrt{\ti})$,
  Eqs. \form{areneq} and \form{match}, and $A_{\mbox{\tiny
      clas}}(\t,\ti)$, the approximation obtained by truncating the
  spectrum at $\mu=\sqrt{\ti}$.  Bottom: Left: Relative error induced
  by truncating the spectrum at $\mu=\sqrt{\ti}$, as a function of
  $\ti$. Right: Difference between two choices of the initial time
  $\ti$ for the fixed-time renormalization scheme, Eqs. \form{areneq}
  and \form{match}. }
\label{aren}
\end{figure}

In Fig. \ref{aren} we compare, for several values of $\ti$,
$A_{\mbox{\tiny ren}}(\t,\mu\!=\!\sqrt{\ti})$ in the fixed-time scheme
with $A_{\mbox{\tiny clas}}(\t,\ti)$, obtained from cutting off
the power spectrum, $P(k,\ti)$ in Fig.~\ref{Pkreg}, at $k=\sqrt{\ti}$.
\footnote{Note that in the linear approximation the truncation of the
  spectrum at $\mu=\sqrt{\t_i}$ is preserved by the time evolution.}.
For $\ti=2$ the maximal difference between them amounts to 2\%,
rapidly decreasing as we increase $\ti$.  A direct comparison between
the values of $A_{\mbox{\tiny ren}}$ and $A_{\mbox{\tiny clas}}$ for
$\ti=2$ is also shown. The goodness of the approximation performed by
truncating the spectrum is clearly evident.  We also study the
dependence of $A_{\mbox{\tiny ren}}$ on the value of $\t_i$ used for
the fixed-time renormalization scheme. We plot the difference between
$A_{\mbox{\tiny ren}}$ defined at $\ti=2$ and $\ti=3$. As it should,
it is of the form $a+b\t$ and it remains very small in all the range
of times we are interested in.

It is easy to estimate the time $\tsb$ when symmetry breaking is expected
and the amplification of modes ceases to take place. 
We can estimate the time of symmetry breaking $\tsb$ by equating
\be\label{phi2}
\langle|\phi|^2(\tsb) \rangle \equiv  M^2 \, N_c  A_{\mbox{\tiny ren}}
(\tsb, \mu=\sqrt{\t_i})= v^2 \equiv N_c \, v_0^2\,.
\ee
We have just described how for $\ti\gsim 2$ a very good approximation
for $A_{\mbox{\tiny ren}}$, in the fixed-time subtraction scheme, is
obtained by just truncating the power spectrum at $\mu^2=\ti$. We can
thus approximate the above expression for the vev by
\be
\langle|\phi|^2\rangle \equiv {  M^2 \,N_c \over 2 \pi^2} 
\int_0^{\sqrt\ti} dk\, k^3\,|f_k|^2 = v^2
\ee
which can be rewritten as
\be\label{pt}
p(\tsb) \simeq \int_0^{\sqrt{\ti}} {dk\over k}\,P(k,\tsb)
= {2\pi^2
\over \lambda \, N_c \,(2V)^{2/3}} \equiv {2\pi^2
\over \lambda_0 \, (2V)^{2/3}} 
\ee
This can be computed exactly using \form{eqfk} but to give an 
analytic estimate of its dependence on the parameters, we will use 
an approximation to the classical power spectrum \form{phi2A}.
In the region of quasi-classical modes it is very well described by,  
\bea\label{Papp}
P_{\rm app}(k,\t) = A(\t)\,k^2\,e^{-B(\t)\,k^2}\,,\hspace{5mm} \\[1mm]
A(\t) = A_0\,\bi^2(\t)\,, \hspace{1cm} B(\t) = 2\sqrt\t-2\,, 
\eea
which can be obtained from $|f_k|^2 \simeq \,|C_1(k)|^2\,|\bi|^2(z)$,
where
\bea\label{A0}{\displaystyle 
k\,|C_1(k)|^2 \simeq {\pi^2 (1/3)^{2/3}\over2\,\Gamma^2(1/3)}\Big(
1 + 2k^2 + {\cal O}(k^4)\Big) \ \simeq \ A_0\,e^{2k^2} } \,,\\
{\displaystyle \exp\Big({4\over3}z^{3/2}\Big) = \exp\Big(
{4\over3}\t^{3/2} - 2\sqrt\t\,k^2 + {\cal O}(k^4)\Big) }\,.
\eea
We have plotted $P_{\rm app} (k,\t)$ together with the exact
spectrum in Fig.~\ref{Pkreg}. We can see that it is an
excellent approximation to the classical power spectrum, in the region of
interest.
Using $P_{\rm app} (k,\t)$ to estimate \form{pt} gives the 
condition
\be\label{pt2}
p(\tsb)  = {2\pi^2
\over\lambda_0  \,(2V)^{2/3}}  \simeq 
\int{dk\over k}\,P_{\rm app}(k,\tsb) = {A(\tsb)\over2B(\tsb)}\,,
\ee
We have evaluated this function $p(\tsb)$ numerically and found an
excellent fit to it, in the range $\t \geq 1$, as, for $N_c=4$, 
\be\label{fit}
\ln\,p(\tsb) = - 3.5 + [8+\tsb^{3.23}]^{1/2}\,,
\ee 
which gives directly the time of 
symmetry breaking in units of $m^{-1}$,
\be\label{tsb}
m t_{\rm sb} = (2V)^{-1/3}\left[\Big(3.5+\ln{2\pi^2\over\lambda_0
(2V)^{2/3}}\Big)^2 - 8\right]^{0.31}\,.
\ee
We can use this compact expression to estimate the time of symmetry
breaking for any coupling $\lambda$ and any inflaton velocity $V$
at the bifurcation. For example, for $\lambda_0=0.11$ and $V = 0.003$,
we find $\tsb = 4.6$ and $m t_{\rm sb} = 25.3$, which agrees very well
with numerical (lattice) simulations performed for those values of
the parameters. 

Note that, as mentioned before, the dependence of $\tsb$ with the 
parameters $\lambda\equiv \lambda_0/N_c$ and $V$ is very mild. 
In the whole range of 
parameters, $\lambda_0\, (2V)^{2/3} \in [10^{-8},\ 1]$, the normalised 
time of symmetry breaking only varies within the range $\tsb = 5 \pm 2$.
Some particular examples can be found in Table~\ref{tnlsb}.

\begin{table}
\caption{The time scales of symmetry breaking and the onset of
the non-linear stage for different model parameters. The coupling
depends on the number of components of the Higgs field
as $\lambda_0= N_c \lambda$.}
\renewcommand{\tabcolsep}{1.7pc} 
\renewcommand{\arraystretch}{1.5} 
\begin{tabular}{|c|c|c|c|c|c|} \hline
$V$ & $\lambda_0$ & $\tnl$ &  $mt_{\rm nl}$ &  $\tsb$ & $mt_{\rm sb}$  \\
\hline
\hline
0.003 & 0.11 & 2.78 & 15.3 & 4.6 & 25.3  \\
\hline
0.003 & 0.01 & 3.76 & 20.7 & 5.2 & 28.5  \\
\hline
0.0003 & 0.001 & 4.01 & 47.5 & 6.0 & 71.3  \\
\hline
0.00002 & 0.0001 & 4.82 & 141. & 6.8 & 200.  \\
\hline
\end{tabular}\label{tnlsb}
\end{table}

\section{Non-linear evolution of the classical system}

In the previous sections we have argued that the first
stages of the quantum evolution of the system (when the
non-linear self-coupling of the Higgs is negligible)
drive the system into a state with highly  populated
low momentum modes. The evolution of this state can be
accounted for by the evolution of a classical
(approximately gaussian) random field. This justifies
the main assumption of this and remaining sections,
namely that the subsequent non-linear dynamics of the
system is determined by the classical evolution of this
field. This evolution is deterministic and
the random character  appears in the initial values of the
field at time $\tau=\tau_i$. These initial conditions are
determined by the exact gaussian quantum evolution of the
system studied in the previous section. Thus, the initial
Higgs field is chosen gaussian, an approximation which can
be tested by probing the sensitivity of our results
to the value of $\tau_i$. As we will see this works
very well within  the appropriate range of initial times.
Some  statistical properties of this initial gaussian random
field can be studied analytically. This is done in section
6.1. These properties extend to times during  which the evolution
is essentially linear and the field remains approximately gaussian.
A full non-perturbative treatment of the dynamics can only be
done by numerical methods. We have actually carried this
step by lattice simulations. This is described in section 6.2
where a full account of the methodology and the checks performed
to show cut-off independence is described.
Results will be presented in the next section

\subsection{Peaks of the Higgs spatial distribution}

The statistics of the Higgs spatial distribution can be determined
from the gaussian fluctuations that are used to build it up. A
detailed description can be found in Ref.~\cite{BBKS} for the case of
the gaussian density field responsible for galaxy formation. In fact,
the spatial distribution and subsequent dynamical behaviour of the
Higgs field at the initial stages of symmetry breaking turns out to be
not that different from that of both the linear and non-linear growth
of the cosmological density field (also built up from the gaussian
random fields of cosmological perturbations), except in the dynamics of
gravitational collapse of the latter.

The fact that the quantum fluctuations of the Higgs give rise to a
classical gaussian random field, allows us to study the statistical
properties of this field in terms of a single function, the two-point
correlation function in Fourier space (i.e. the power spectrum), whose
approximate expression can be found in \form{Papp}. This quantity
allows the computation of several related quantities that characterise
a gaussian random field, e.g. the spatial correlation function, the
density of peaks above a certain threshold, the shape of the highest
peaks, etc.

The first quantity that we can compute is the spatial correlation
function, defined as the two-point correlation function 
between two points separated by a distance $r$,
\bea\nonumber
\xi(r,\t) &\equiv& \left\langle\phi(r,\t)\phi(0,\t)\right
\rangle = { M^2\, N_c \over2\pi^2}
\int_0^\infty {dk\over k}\,P_{\rm app}(k,\t)\,{\sin\,kr\over kr}\\[1mm]
&\simeq& { M^2\, N_c \over2\pi^2} \nonumber
{A(\t)\over r}\int_0^\infty dk\,e^{-B(\t)k^2}\,\sin\,kr \\[1mm]
&=&{M^2\, N_c \over2\pi^2}
{A(\t)\sqrt\pi\over r\,2B^{1/2}(\t)}\,\exp\Big(\!-{r^2\over4B(\t)}
\Big)\,{\rm erfi}\Big({r\over2B^{1/2}(\t)}\Big)\,,\label{corr}
\eea
where $\ {\rm erfi}(x)$ is the imaginary error function~\cite{AS}.
This correlation function determines the average size of the lumps,
$\xi_0$, 
\be
\xi_0 \simeq 2B^{1/2}(\t) \simeq 2\sqrt{2\sqrt\t-2}\,,
\ee 
as a function of normalised time $\t$. Note that the time dependence
of the correlation length is different than for a quench symmetry
breaking. While in the latter case, the correlation length grows like
$\xi_0 \sim 2\sqrt\t$, in our case, it grows like $\xi_0 \sim
2\sqrt2\,\t^{1/4}$ for ``large'' $\t$ (still in the linear
regime). This introduces some slight differences in the behaviour of
the field at symmetry breaking.

We can then compute from \form{corr} the time-dependent dispersion
\be\label{phirms}
\sigma(\t) \equiv \xi^{1/2}(0,\t) = \phi_{\rm rms}(\t)=
{ M\, \sqrt{N_c}\over\sqrt2\,\pi}\,p^{1/2}(\t)\,,
\ee
which is nothing but the root mean square value of the Higgs field.

Another quantity which is very useful to characterise the field
distribution is the number density of peaks of the field above a
certain threshold $\phi_c$, see Ref.~\cite{BBKS,HGW}
\bea
n_{\rm peak}(\t) = {1\over4\pi^2}\left({-\xi''(0,\t)\over\xi(0,\t)}
\right)^{3/2}(\nu^2-1)\,\exp(-\nu^2/2)\,,\\[2mm]
{-\xi''(0,\t)\over\xi(0,\t)} = {\langle k^2\rangle\over3} =
{\int (dk/k)P(k,\t)k^2\over3\int (dk/k)P(k,\t)}\,,\hspace{1cm}
\eea
where $\nu = {\phi_c\over\sigma(\t)} > 1$. In our case, the number density 
of high peaks is given by 
\bea
n_{\rm peak}(\t) &=& {2\over3\sqrt3\,\pi^2}\,{1\over r_0^3(\t)}\,
(\nu^2-1)\,\exp(-\nu^2/2)\,,\\[2mm]
\nu(\t) &=& {\phi_c\over v}\,{\sqrt2\,\pi\over
(2V)^{1/3}\sqrt\lambda_0}\,p^{-1/2}(\t)\,.
\eea
We have evaluated this function for the parameters $\lambda=0.11/4$ and
$V=0.003$, at various times and compared with our lattice simulations
(for different volumes $V=(2\pi/p_{\rm min})^3$ and lattice spacings
$a$). The results are very encouraging. If we multiply this density of
peaks by the actual volume of the simulations, we find indeed just a few
peaks above e.g. $\phi_c = 0.02 v$, at the time of symmetry breaking.

\begin{figure}
\includegraphics[width=10cm,angle=-90]{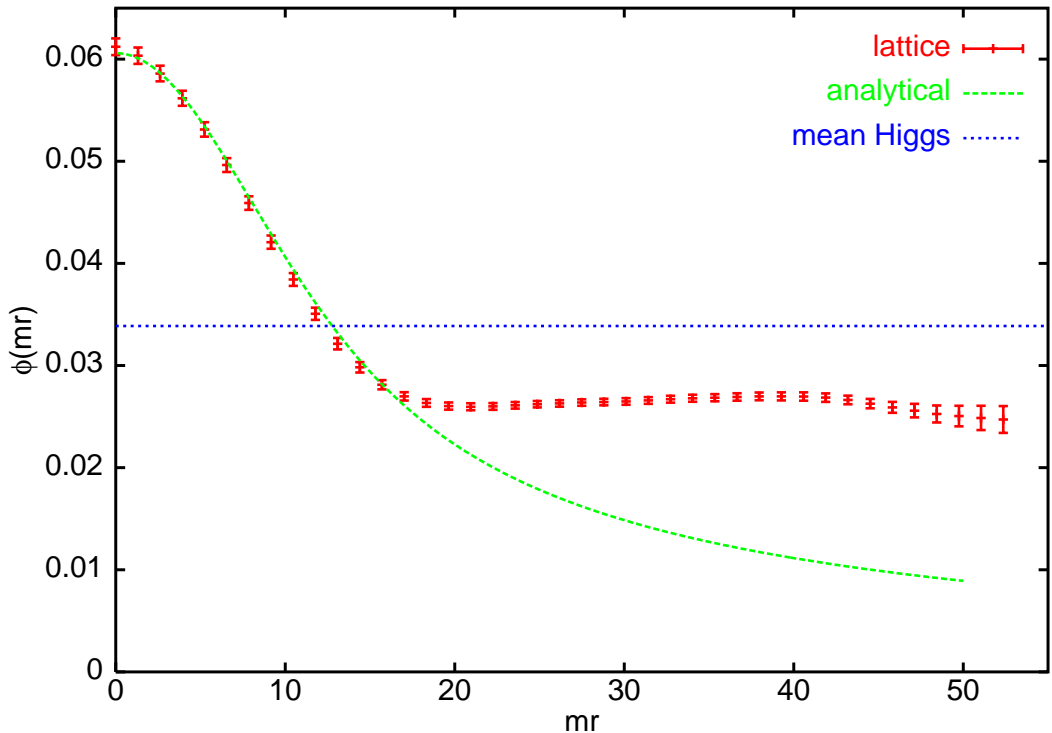}
\caption{The
  radial profile of the Higgs peak for $\lambda=0.11/4$ and
  $V=0.003$, at time $\t=2.54$, corresponding to $m t = 14$, obtained
  with our lattice simulation (with error bars, from averaging over
  several realizations), and compared with the analytical result
  \form{peak}.  We have also included the rms Higgs value
  \form{phirms} at that time. Note that we are still in the linear
  regime, where \form{peak} gives a very good approximation. The
  higher tail corresponds to an averaging out of several lower peaks.}
\label{lump}
\end{figure}

In fact, we can compute not only the probability per unit volume to
find a peak in the distribution of the Higgs field, but also their
radial profile~\cite{BBKS},
\bea\nonumber
\rho(r,\t) &=& {M\, \sqrt{N_c}\over\sqrt2\,\pi}
\int_0^\infty {dk\over k}\,P^{1/2}_{\rm reg}(k,\t)\,
{\sin\,kr\over kr}\\[1mm] \nonumber
&\simeq& { M\, \sqrt{N_c} A^{1/2}(\t) \over\sqrt2\,\pi \,r} 
\int_0^\infty {dk\over k}\,e^{-{B(\t)\over2}k^2}\,\sin\,kr \\[1mm]
&=& { M\, \sqrt{N_c} A^{1/2}(\t) \over 2 \sqrt2\, r}
\,{\rm erf}\Big({r\over\sqrt{2B(\t)}}\Big)\,,
\label{peak}
\eea
where $\ {\rm erf}(x)$ is the error function~\cite{AS}. We have plotted
this profile function in terms of the radial coordinate, together with
the lattice results in Fig.~\ref{lump}, for $\lambda=0.11/4$ and
$V=0.003$, at time $\t=2.54$, corresponding to $m t = 14$, well
before symmetry breaking, which occurs at $m t_{\rm sb} \simeq 26$. 

\subsection{Lattice simulations}\label{lattice}

The previous analysis falls short of addressing the most important
aspects of symmetry breaking after hybrid inflation since the main
effect is non-perturbative, see Ref.~\cite{GBKLT}. 
As discussed we will incorporate these effects by performing
classical real-time numerical simulations in the lattice.
Generically this classical approximation would fail 
to reproduce the relevant physics but we have just argued that this is 
indeed the correct approximation for the infrared modes of the Higgs 
at the time of symmetry breaking. 

The usual procedure \cite{FT} \footnote{ See also 
\cite{TK,PR,GBKLT,RSC,GBRM,CPR}, where the same or analogous initial
conditions have been used.} is to take as initial conditions
for the lattice simulations gaussian random fields given by the
distribution \form{wigner2} with vacuum initial amplitudes 
corresponding to Eqs. (\ref{fkw}), (\ref{gkw}). We would like to stress
here that the correct description of the quantum linear system in
terms of a gaussian random field requires the use of two independent
gaussian variables, as indicated in \form{wigner2}. One of them, $y$ in 
\form{wigner2},  describes field fluctuations with dispersion $|f_k|^2$
and a random phase. The other, $\bar p$, with dispersion 
$(4\,|f_k(\t)|^2)^{-1}$ and a random phase,  allows to define the conjugate 
momentum through, see Appendix B, 
\be \label{moment}
 p = \bar p + {F_k(\t) \over |f_k(\t)|^2} \, y \,.
\ee

Notice that this prescription is valid in order to give initial
conditions at {\em any} time during the evolution before
non-linearities set in.  In particular, as described in section
\ref{nonlin}, we propose to take as starting point for the lattice
simulations the above gaussian ensemble at a fixed time sufficiently
advanced to guarantee that a large fraction of modes have become
classical, but well before the time when non linearities become
relevant, in a similar spirit as that in Ref.~\cite{Smit}. 
This has the advantage of allowing a clear separation
between infrared (classical) modes which evolve classically and
ultraviolet (quantum ) modes that will be absorbed in the
renormalization of the constants of the theory. From the previous
analysis, see Fig. \ref{Fk1} and the discussion after Eq. (\ref{tsb}),
a good choice for the matching time in a wide range of model parameters
seems to be $\tau_i \simeq 2$. See the discussion in the next section
about the onset of the non-linear regime.

Therefore, we propose the following as initial conditions in our lattice 
simulations. At a fixed time $\tau_i$ previous to symmetry breaking: 

\begin{itemize}

\item
Put to zero all the modes that have not become classical at
$\tau_i$. This includes all the modes of the inflaton but the
homogeneous zero mode, and all large momentum modes of the Higgs with
$k> \sqrt{\t_i}$ (replacing the hard cutoff at $\sqrt{\t_i}$ by the
approximate power spectrum in \form{Papp}, which strongly dumps
ultraviolet modes, does not significantly change the results even at a
quantitative level).

\item
Set the homogeneous zero mode of the inflaton to $\chi/\chi_c= 1-Vm\,t_i$ 
with conjugate momentum $\dot \chi/\chi_c= -Vm$. 

\item For the Higgs fluctuations, each Fourier component, with momentum
$|k|\le \sqrt{\ti}$,  has an 
amplitude $|\phi_k|$ randomly generated according to    
the Rayleigh distribution:
\be
 P(|\phi_k|)\,d|\phi_k|\,d\theta_k =
\exp\Big(\!-{|\phi_k|^2\over\sigma_k^2}
\Big)\,{d|\phi_k|^2\over\sigma_k^2}\,{d\theta_k\over2\pi}\,,
\ee
 with dispersion given by $\sigma_k^2 = |f_k|^2 =
k^{-3}\,P(k,\tau_i)$, and a uniform random phase
$\theta_k \in [0,2\pi]$. The conjugate momentum $g_k = i\phi'_k$ is
uniquely determined once $\phi_k$ is known, through the relation
\be
\phi_k ' = {F_k(\tau_i) \over |f_k(\tau_i)|^2 }\  \phi_k,
\ee
with $f_k(\tau_i)$ and $F_k(\tau_i)$ given by Eqs.~(\ref{eqfk})
and (\ref{Fk}), respectively, at $\tau=\tau_i$.
This corresponds to the classical limit of \form{moment}, an approximation 
that is well justified for $\ti \gsim 2$, see Appendix B.

\item
Take the masses and couplings used in the simulation as the physical 
renormalised ones in the fixed-time subtraction prescription. 
\end{itemize}

\begin{figure}\includegraphics[width=.67\textwidth,angle=-90]{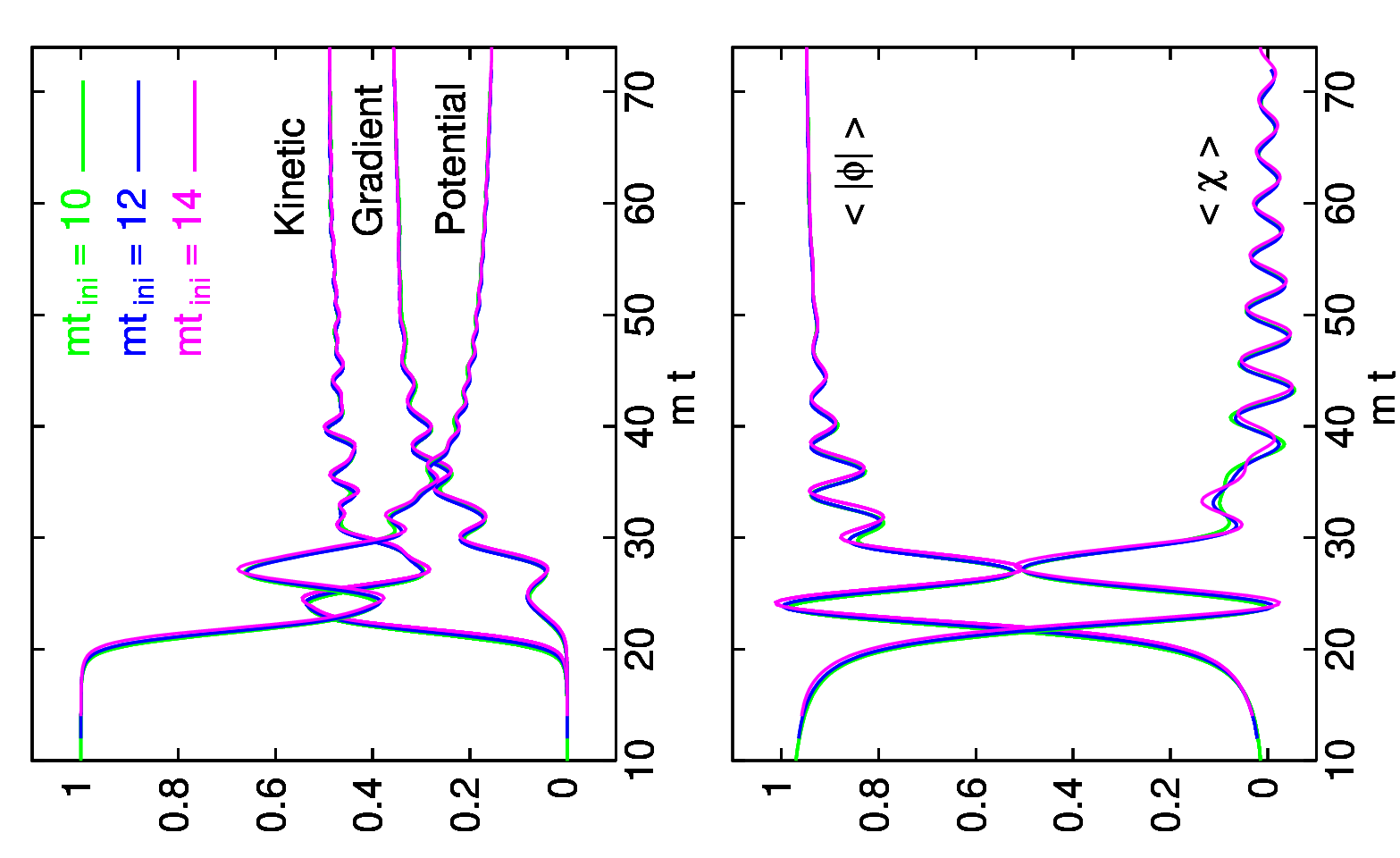}
\includegraphics[width=.67\textwidth,angle=-90]{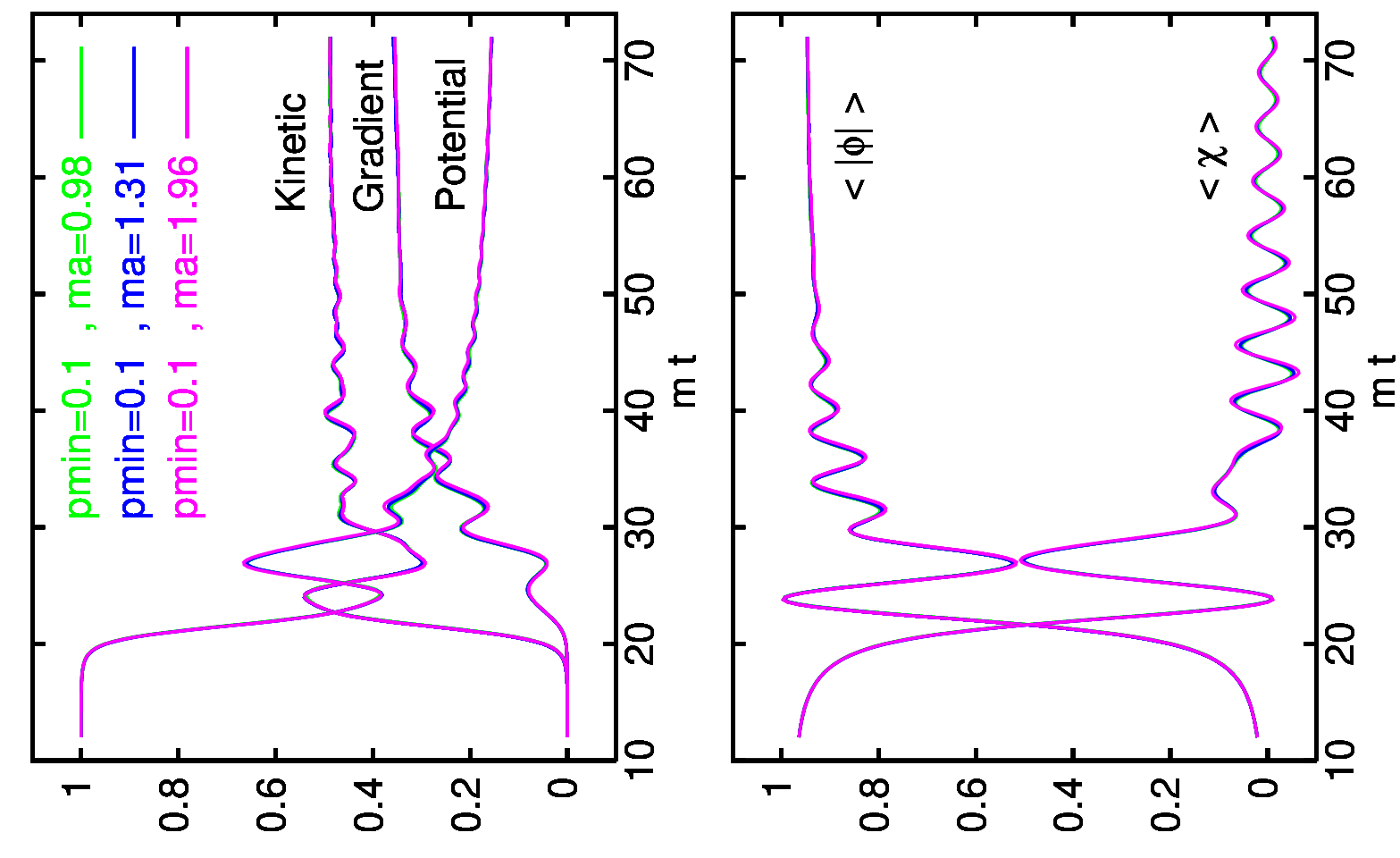}
\caption{The time evolution of $\langle |\phi|\rangle$, $\langle \chi \rangle$
and energies (normalised to the initial one) for $\lambda
\!=\!0.11/4$ and $V\!=\!0.003$, obtained with our lattice simulation.
Left: for different choices of $mt_i$, the time for matching the
quantum evolution to the classical lattice
simulations. Right: for different values of the lattice spacing
$ma=0.98, 1.31,1.96$.}
\label{Ftini}
\end{figure}

As long as the time chosen for initialisation is sufficiently advanced 
that a large fraction of modes have become classical, we hope that most of 
the phy\-sics responsible for symmetry breaking will be included in 
the simulations. How advanced it has to be in a concrete realization 
can be tested by studying in which range the time evolution is insensitive 
to the choice of $\tau_i$. This provides also a check 
of the validity of our approach. The result of such a test 
is presented in Fig. \ref{Ftini}. We compare 
the time evolution of  $\langle |\phi|\rangle$, $\langle \chi \rangle$
and the average kinetic, gradient and potential energies
obtained from setting the initial conditions at $mt_i=10,12,14$ 
($\tau_i= 1.81, 2.18, 2.54$) for $\lambda =0.11/4$ and $V=0.003$. 
The agreement is excellent, corroborating our estimate that for 
$\tau_i\simeq 2$ all the basic relevant modes driving symmetry 
breaking have already become classical and thereafter the evolution
is well described by our lattice classical simulations.

All the lattice results presented in this paper have been obtained for
a SU(2) Higgs doublet coupled to the inflaton with coupling
$g^2=2\lambda =0.22/4$, and inflaton velocity $V=0.003$.  Due to the
finite volume, the momentum $k$ is discretised in units of a minimal
momentum given by $p_{\rm min} = 2\pi /L$, with $L=Na$, where $N$ is
the number of lattice points, and $a$ the lattice spacing.  Our
simulations have been performed in lattices of sizes $32^3$, $48^3$
and $64^3$ with physical volumes determined by $p_{\rm min}=0.1m, \
0.075m$ and $0.05m$ and lattice spacings varying from $ma\simeq 1$ to
$ma\simeq 2$.  The choice of lattice volumes and lattice spacings has
been performed such as to avoid lattice spacing and finite volume
dependence of the lattice results.  Notice that the minimal momentum
has to be small enough that a sufficiently large number of classical
momenta with $k\lsim 1$ is taken into account. We have found that for
$p_{\rm min} \leq 0.1m$ this is indeed the case and no significant
volume dependence is observed.

A further essential test of our approach is that it succeeds in taming
ultraviolet divergences. On the lattice there is a maximal momentum
determined by the lattice cutoff through $p_{\rm max} = 2\pi / a$.
Naturally, re-scattering and back-reaction will populate the high
momentum modes at and after symmetry breaking This is certainly a
physical effect but if the lattice cutoff is not chosen large enough
population of the high momentum modes is artificially induced by
cutoff effects. The lattice cutoff should then be chosen such as to
avoid that this takes place before the relevant dynamics of symmetry
breaking. A reasonable value for our choice of parameters is $ma < 2$,
as can be seen from Fig. \ref{Ftini}, where we compare the time
evolution of $\langle |\phi|\rangle$, $\langle \chi \rangle$ and
average energies for several values of the lattice cutoff $ma=0.98,
1.31,1.96$. No significant lattice spacing dependence is observed
here, while it becomes clearly appreciable for $ma>3$.  Details of the
simulations and further results will be presented elsewhere.  These
lattice simulations will allow us to test the next stage, the
non-linear approach to symmetry breaking.

\section{Results of the non-linear analysis and ``bubble'' formation}

\begin{figure}
\includegraphics[width=14cm,angle=0]{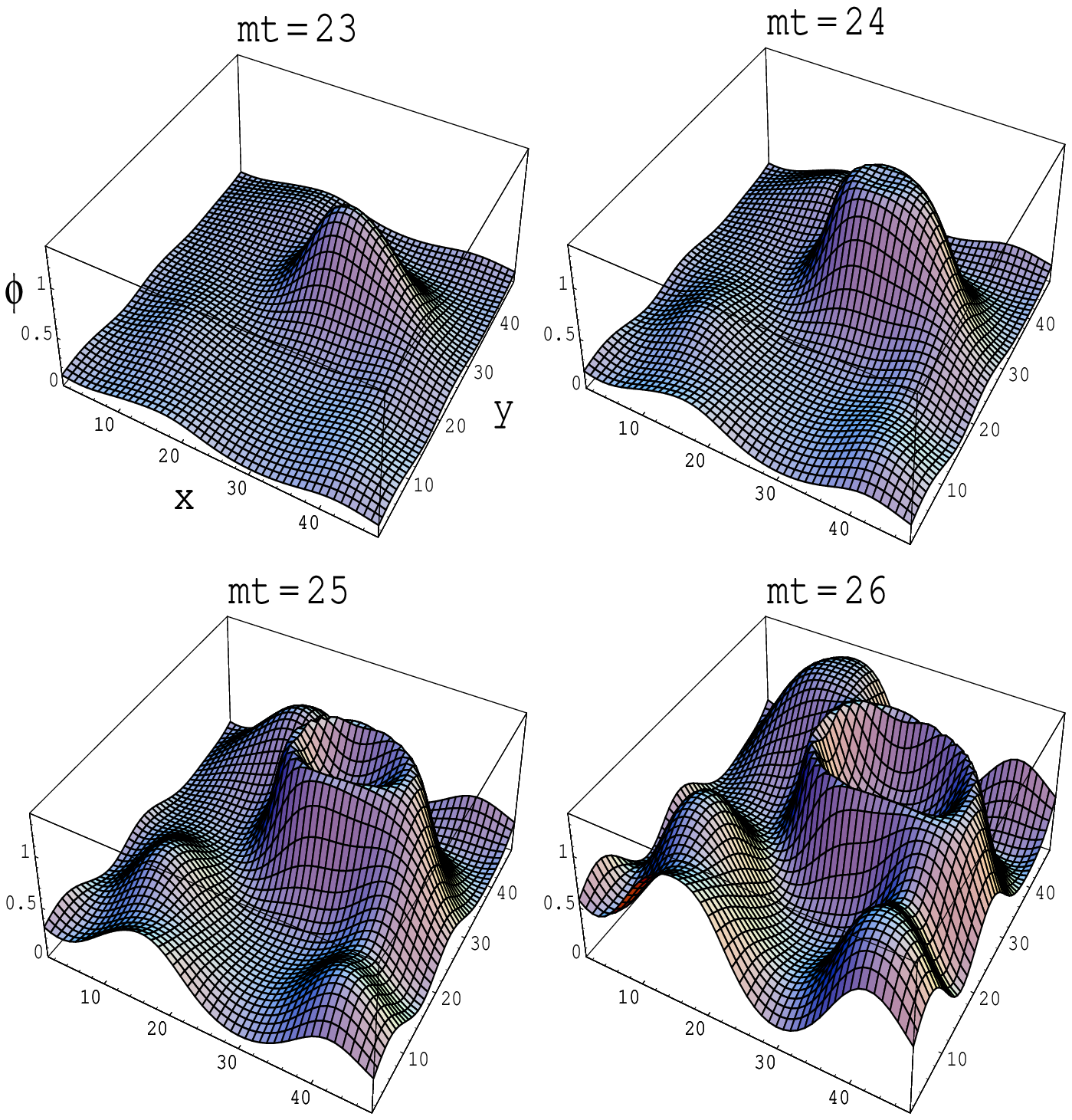}
\caption{Snapshots of the growth of the Higgs peak in a full non-linear
   lattice simulation, for $\lambda=0.11/4$ and $V=0.003$. Plotted is the
   value of the Higgs amplitude
   $\phi$ in the plane $(x,\ y)$, where the $z$-coordinate is that of
   the highest peak.  Note that several peaks appear in the simulation.
   Here we show the first stages of the evolution, where the highest
   peak invaginates and forms what we call the ``bubble''.}
\label{fig2duno}
\end{figure}

\begin{figure}
\includegraphics[width=14cm,angle=0] {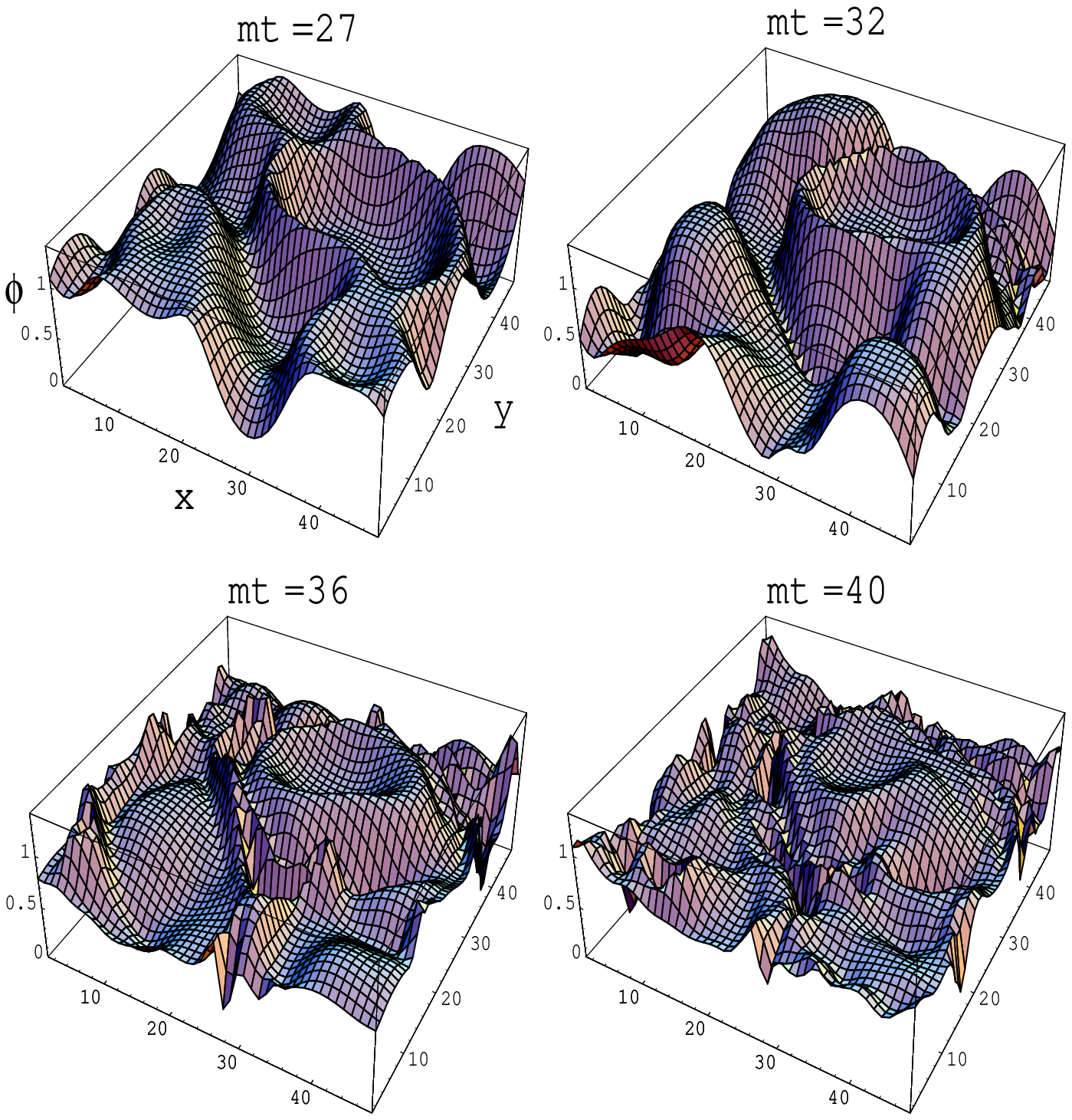}
\caption{Same as
   in Fig.~\ref{fig2duno}. Here we show the late stages, in which
   gradients arise from collisions of bubbles and the symmetry is
   broken, i.e. $\phi\simeq1$.}
\label{fig2ddos}
\end{figure}

In the previous sections we have developed a formalism to describe the
linear growth of the Higgs quantum fluctuations and their conversion
into a classical gaussian random field. As we have argued, in the
linear regime there is only the homogeneous mode of the inflaton,
$\langle\chi/\chi_c\rangle = 1 - Vm(t-t_c)$, which induces a negative
mass squared \form{masst} for the Higgs, and thus its spinodal
instability towards the true vacuum. The quick growth of the quantum
fluctuations generates a gaussian random field with correlation
function \form{corr}, and a rms field value, see \form{phirms},
\be\label{phit}
\phi(\t) \equiv {\phi_{\rm rms}\over v} = {(2V)^{1/3} \sqrt{\lambda N_c}
\over\sqrt2\,\pi}\,p^{1/2}(\t)\,,
\ee
where $p(\t)$ is given by \form{pt}. Eventually, the mean field $\phi$
will become large and will approach the vev of its potential, thus
breaking the symmetry. Before that happens, its coupling to the
inflaton will induce a back-reaction on the homogeneous mode of the
inflaton, $\chi$, which will start to deviate from the linear regime
described above. At this stage the 
non-perturbative evolution can be studied by numerically solving the 
coupled classical equations of motion for the inflaton and Higgs:
\bea
\label{homeq} {\displaystyle
\ddot\phi^a(\x,t) - \nabla^2\phi^a(\x,t) +
(|\phi|^2(\x,t) + \chi^2(\x,t) - 1)\,\phi^a (\x,t) = 0\,,}\\[1mm]
{\displaystyle
\ddot\chi(\x,t) - \nabla^2 \chi(\x,t) +
{g^2\over\lambda}|\phi|^2(\x,t) \,\chi(\x,t)  = 0\,,}\hspace{2.5cm}
\eea
with gaussian initial conditions as described in the previous sections.
In this section we will present the results of our numerical simulations
and give an approximate analytic understanding of
how symmetry breaking takes place.

Although the initial conditions are random, as a result of the
non-linear dynamics many of the {\em qualitative} features of the
evolution are fairly universal, although {\em quantitatively}
different configurations differ by small shifts in the origin of times
as well as spatially random positions for the center of the peaks.
Therefore, we prefer to illustrate our analytic formulas by comparing
with the results of a {\em typical} lattice configuration, e.g. the
one displayed in Figs.~\ref{Ftini}-\ref{fig2ddos}.

\begin{figure}
\includegraphics[width=.32\textwidth,angle=-90]{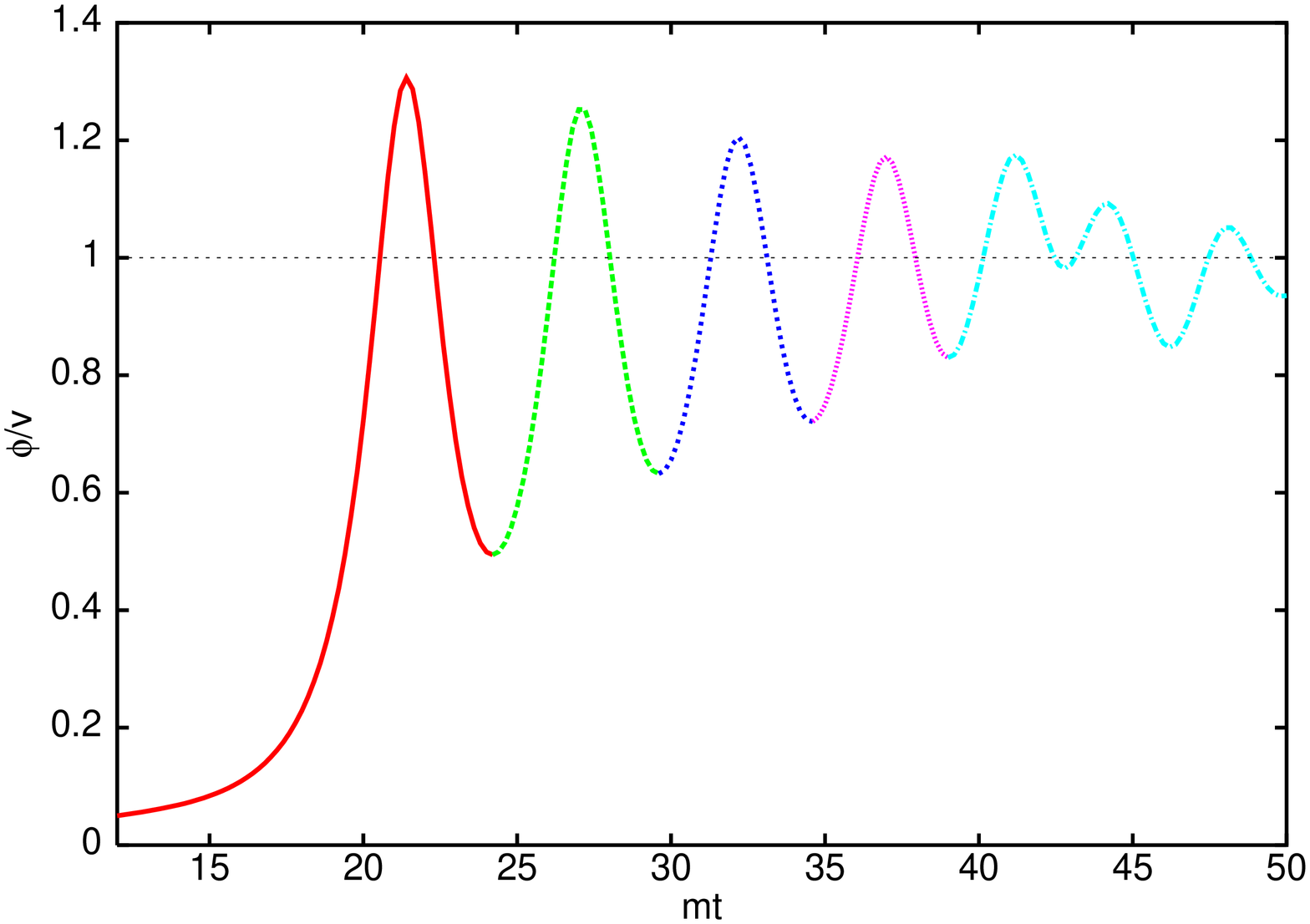}
\includegraphics[width=.32\textwidth,angle=-90]{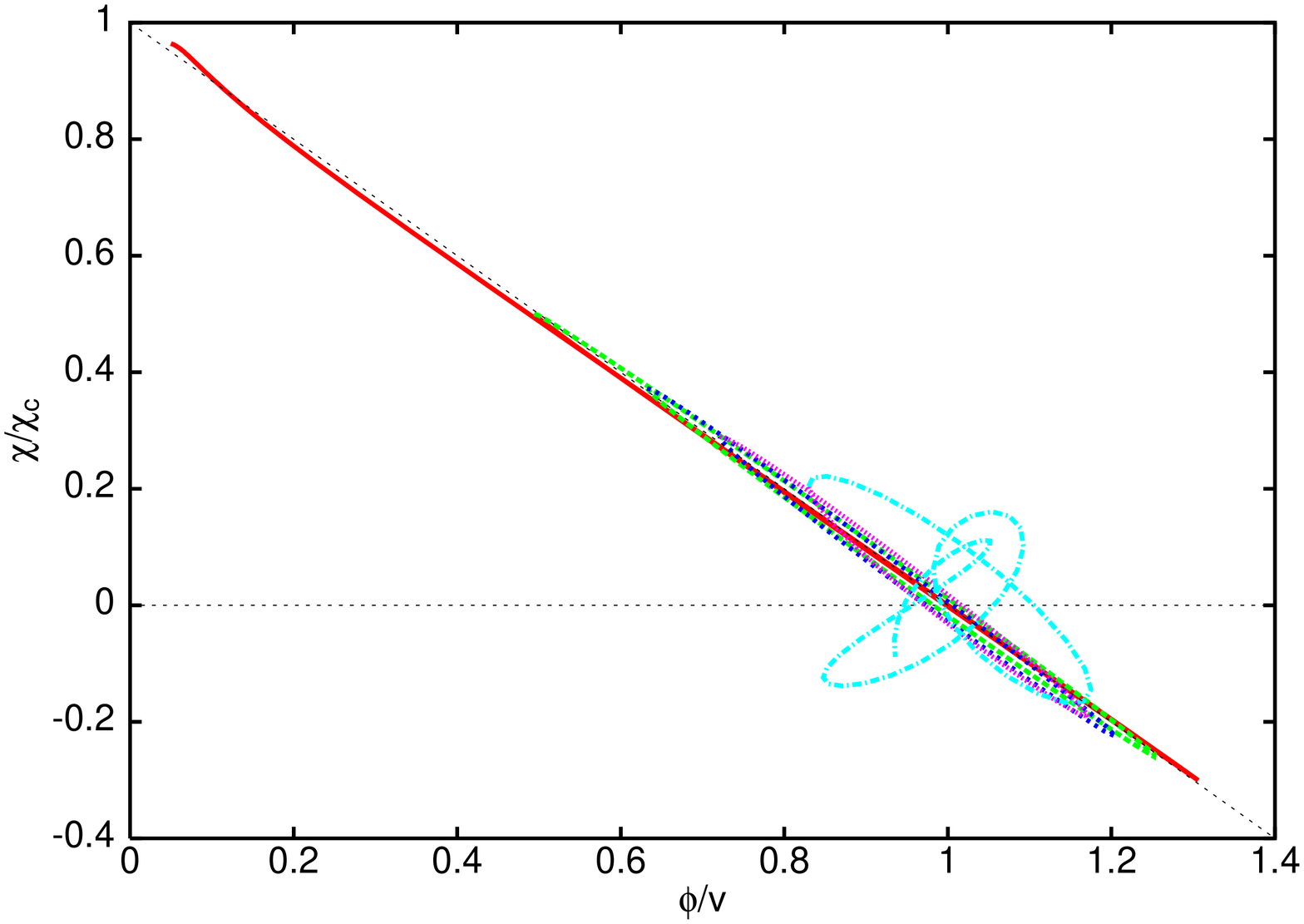}
\caption{ Left: The time evolution of the modulus of the Higgs, at the
location of the highest Higgs peak.  Plotted is the Higgs modulus
$\phi/v$ as a function of time.  Note the effect of bubble collisions on
the Higgs oscillations after $mt=40$.  Right: Collinear evolution of the
inflaton and the Higgs at the location of the highest Higgs peak. Note
that the inflaton and Higgs satisfy $\phi = 1 - \chi$ to very good
accuracy, until rather late, when ``bubbles'' start to collide.}
\label{topbub1}
\end{figure}

Symmetry breaking in our model is not at all a homogeneous process.
Already in the linear regime, the Higgs field evolves by developing
lumps in space that grow with time, see \form{peak} and
Fig.~\ref{lump}.  The classical evolution of the Higgs' lumps, once
non-linearities become relevant, can be followed in Figures
\ref{fig2duno} and \ref{fig2ddos} where we show some snapshots of the
growth of the Higgs' peaks from the first stages of the evolution,
$mt=23$, till $mt=40$ above which full symmetry breaking takes place
and the mean Higgs field approaches the vev. As can be seen from the
figures, the peak of the largest Higgs' lump is the first to break the
symmetry, i.e. to reach $|\phi|=v$, and soon after the center of the
lump invaginates, creating an approximately spherically symmetric
``bubble'', with ``ridges'' that remain above $|\phi|=v$.  Finally,
neighbouring bubbles collide and the symmetry gets fully broken
through the generation of higher momentum modes.  In
Fig. \ref{topbub1} we show the behaviour of $|\phi(\x,t)|$ at the
center of the highest Higgs lump. It oscillates around $|\phi|=v$ with
an amplitude that is dumped in time. Oscillations remain coherent
giving rise to concentric bubbles, until the time when bubble
collisions break the symmetry.

It is possible to get an analytic understanding on how this non-linear
process takes place before bubbles start to collide.  For the problem
we are considering, we can rewrite the components of the Higgs field
as $\phi^a \equiv \phi\, \hat n_a$ (we will use from now on the symbol
$\phi$ to denote the modulus of the Higgs) while $\Omega=\hat
n\cdot\sigma \in$ SU(2) is an element of the gauge group, with
$\sigma=(\uni,\ i \vec{\tau})$ with $\tau_a$ the Pauli matrices. With
this the equations of motion for the coupled inflaton Higgs field can
be rewritten as
\bea\label{homeqsph} &{\displaystyle
\ddot\phi(\x,t) - \nabla^2 \phi(\x,t) +
(\phi^2 + \chi^2 - 1)\,\phi  - \,|\partial_\mu \hat n |^2 \, 
\phi = 0\,,}\\[1mm] \label{constr} &{\displaystyle
\partial^\mu(\phi^2\,\partial_\mu\hat n) = 
- \hat n  \, \, \phi^2 \, |\partial_\mu\hat n |^2}\,,\\ 
\label{homeq2sph} &{\displaystyle
\ddot\chi(\x,t) - \nabla^2\chi(\x,t) +
{g^2\over\lambda}\phi^2\,\chi = 0\,,}
\eea
where dots and $\nabla$ denote derivatives with respect to $m t$ and $m
\x$ respectively, and the homogeneous modes have been normalised to
their vevs, $\phi/v \to \phi$ and $\chi/\chi_c \to \chi$.

\begin{figure}
\includegraphics[width=10.5cm,angle=-90] {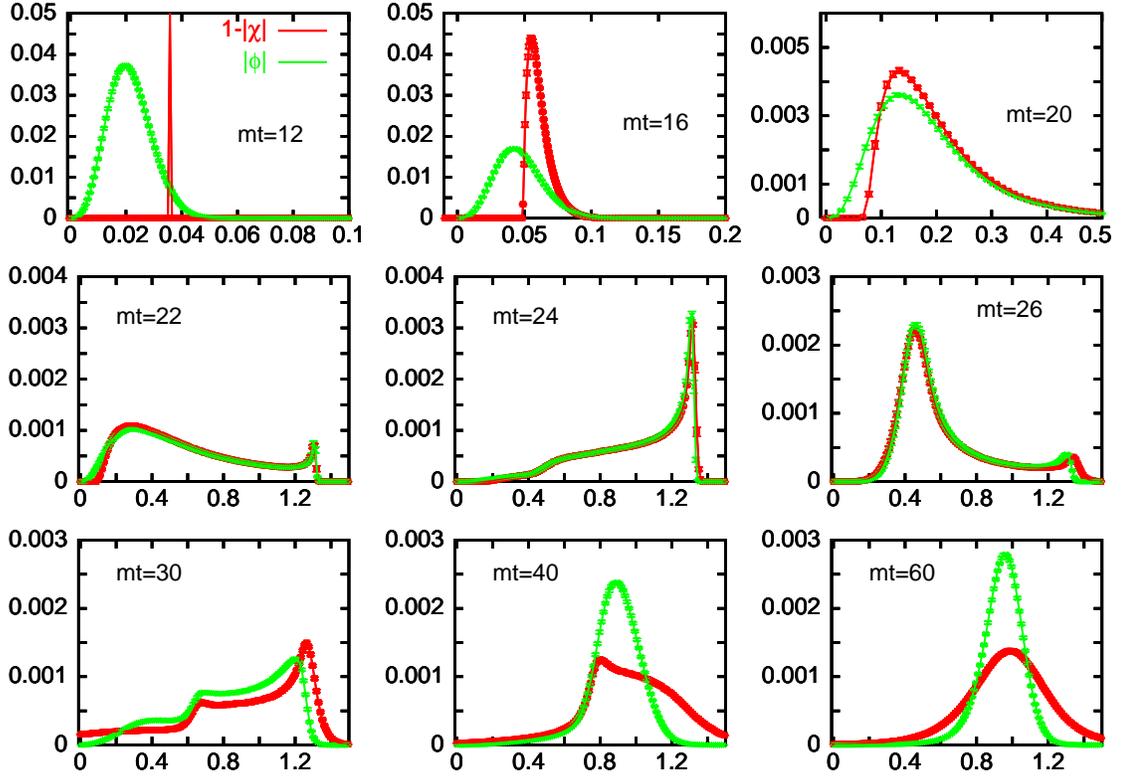}
\caption{The histogram
  of field values for the Higgs modulus $\phi$ and the co-inflaton $1
  - \chi$. Note that soon after the initial condition and the
  subsequent non-linear evolution, the two coincide very precisely.
 Note that a few oscillations can be observed during symmetry breaking,
i.e. during times $mt=22-30$.}
\label{hist}
\end{figure}

We can take advantage of the fact that, for $g^2=2\lambda$, a solution
to the set of coupled equations of motion is given by
$\phi(\x,t)=1-\chi(\x,t)$ \cite{BGK,GBKLT} and
$\partial^\mu(\phi^2\,\partial_\mu\hat n) =0$.  Numerical results
corroborate that this is very approximately the solution soon after
non-linearities set in. In Figure \ref{topbub1} we show, for our model
with parameters $V=0.003$ and $\lambda=0.11/4$, $\phi$ versus $1-\chi$
at the location of the highest Higgs lump.  Comparing with Figure
\ref{topbub1} we can follow how the Higgs and inflaton evolve
colinearly during all the time of coherent oscillations of the peak.
In Fig.~\ref{hist} we show the distribution of field values as a
function of time, in the time interval between $mt=12$ and $mt=60$,
where most of the action takes place. During most of the non-linear
initial stage, through symmetry breaking and until bubbles collide we
have: $\phi(\x,t)=1-\chi(\x,t), \ \forall \x$ .

\begin{figure}
\includegraphics[width=10cm,angle=-90]{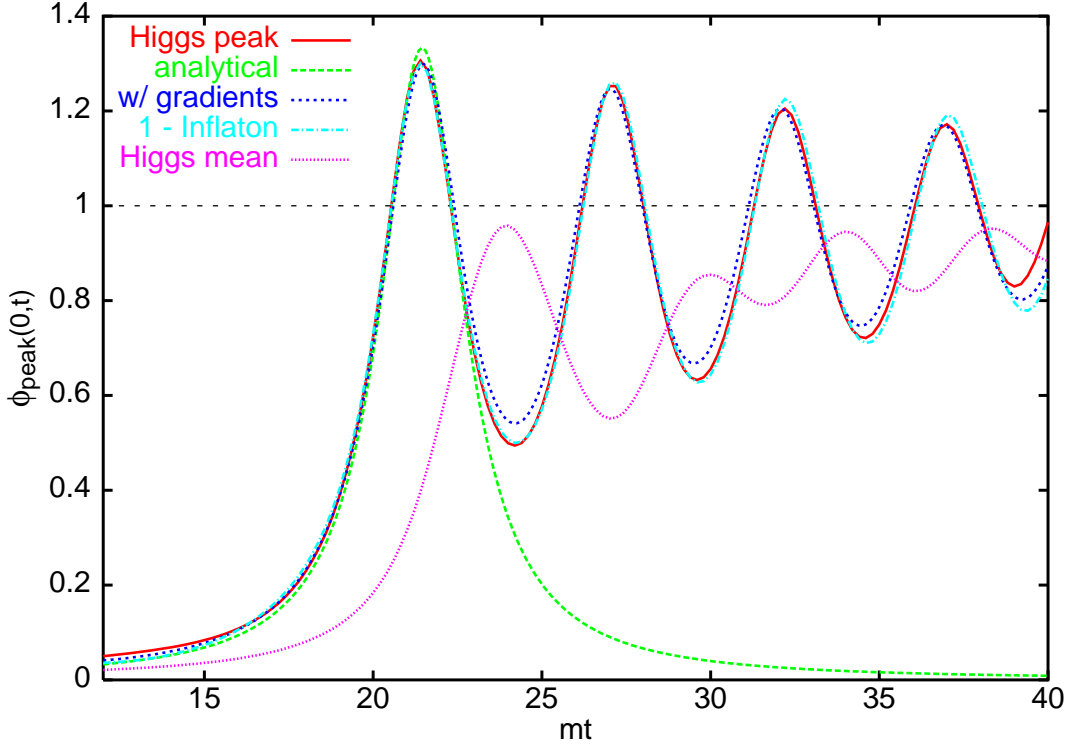}
\caption{The time evolution of the Higgs peak ($r=0$) and the Higgs rms value,
  obtained with our lattice simulations, as compared with the
  analytical result \form{nlsol}, and the numerical solution of
  equation~\form{partialrt}, which includes the gradient terms. Also
  shown is the comparison between the Higgs and the inflaton
  evolution, i.e. $\phi(t)$ and $1-\chi(t)$.}
\label{peaktime}
\end{figure}

\begin{figure}
\includegraphics[width=10cm,angle=-90]{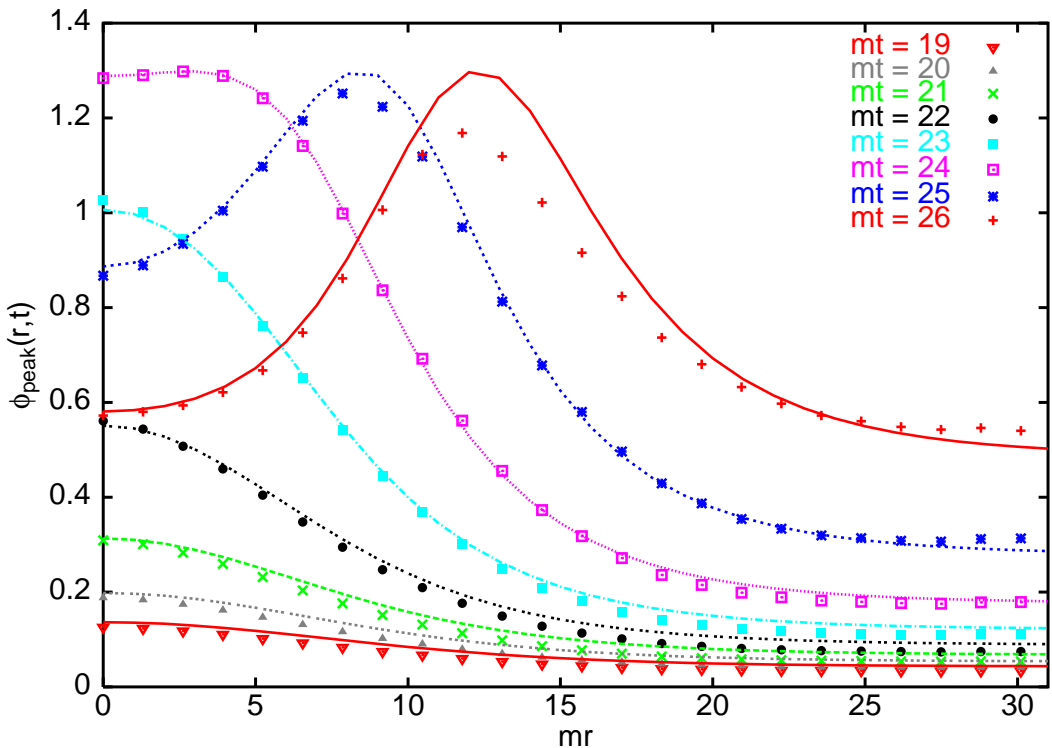}
\caption{ The
  time evolution of the Higgs radial profile around the highest peak,
  obtained with our lattice simulation (points), as compared with the
  numerical solution of the partial differential equation (lines). It is
  surprising how well the formation of the bubble is reproduced with
  the simple assumption of homogeneity around the peak of the bubble.
  Of course, the peak solution does not take into account the presence
  of secondary bubbles, that appear in the lattice simulation at $r
  \sim 30$.}
\label{bubbles}
\end{figure}

During the time that inflaton and Higgs evolve colinearly, the system can 
be seen as that of a {\rm single} field with a modified potential 
$\bar V(\phi)$, with the minimum at $\phi=1$.
\bea\nonumber
{\cal S} &=& {3\over2}\,\int d^3\x\,dt\left[
\half(\partial_\mu\phi)^2 - \bar V(\phi)\right]\,,\\
&&\bar V(\phi) = {1\over6}\Big(1 - 4\phi^3 + 3\phi^4\Big)\,.
\eea
The equation of motion of the scalar field $\phi$ becomes a non-linear
partial differential equation 
\be\label{partial}
\ddot\phi(\x,t) - \nabla^2 \phi(\x,t) -
2\phi^2(\x,t) + 2\phi^3(\x,t) = 0\,.
\ee
If the gradient terms are much smaller than the non-linear ones, 
we can as a first approximation neglect them leading to
\be\label{homo}
\ddot\phi(t) - 2\phi^2(1 - \phi) = 0\,,
\ee
which leads to a conserved energy  $E= E_0 + {1\over 6}$ with 
\be
E_0 \equiv  \half \Big( \dot \phi^2(t) - \phi^3  ({4 \over 3} -\phi) \Big ) \,.
\ee
A solution with $E_0=0$, a very good approximation taking into
account that initially both the field and its derivative are very small,
is given by
\be
\label{nlsol}
\phi(t) = 1 - \chi(t) = {12\over 9 + 4[mt - m t_{\rm max}]^2}\,,
\ee
with $t_{\rm max}$ the time at which the field reaches its
maximum value: $\phi(t_{\rm max}) = {4 \over 3}$. This time
can be rewritten in terms of the value of the field at any other time 
$t_{\rm nl}$ through
\be
m t_{\rm max} = m t_{\rm nl} + \sqrt{{3\over\phi_0} - {9\over4}}\,.
\ee
In particular, at every point $\x$, we can take $\phi_0$ as the
``initial'' value of the Higgs field. This is given by the profile of
the lump in the linear approximation, Eq. \form{peak}, at a time
$t_{\rm nl}$ at which the evolution becomes non-linear and we can no
longer ignore its higher order interactions.  In Fig.~\ref{peaktime}
we show again the non-linear growth of the Higgs field at the top of
the largest peak in the simulation, and compare it with the analytical
solution \form{nlsol}. The agreement is very good during the first
oscillation although \form{nlsol} cannot reproduce the subsequent
ones.  At these stage we can already understand how the spherical
bubbles arise. Take the spherically symmetric peak profile \form{peak}
at the non-linear time $t_{\rm nl}$ and let each point $\x$ evolve
like \form{nlsol}. Points with higher value of $\phi_0(t_{\rm nl})$
will reach first the maximum value ($\phi_{\rm max}={4 \over 3}$) and
then decrease. This generates a spherical wave that propagates from
the center of the lump to infinity. The production of bubbles
associated with symmetry breaking were first described in
Ref.~\cite{GBKLT} for the model $\lambda\phi^3$, which is analogous to
our reduced model. The subsequent evolution is of course different,
due to the presence of the inflaton field.

We can evaluate the non-linear time, $t_{\rm nl}$,  by equating 
$$\chi(\tnl) = 1 - \half(2V)^{2/3}\tnl = 1 - \phi(\tnl)\,.$$  
Using \form{phit} and \form{fit}, we can find the
nonlinear time $\tnl$ as the solution of the transcendental equation
\be
\t = \left[\Big(3.5 + 2\,\ln{\pi(2V)^{1/3}\t\over\sqrt{2\lambda N_c}}
\Big)^2 - 8\right]^{0.31}\,.
\ee
For the values of parameters chosen, $\lambda= 0.11/4$ and $V =
0.003$, we find $mt_{\rm nl} = 15.3$ and $\phi_0 = 0.1$. That is, soon
after the Higgs field becomes non-linear, it ceases to grow
exponentially like \form{phit}, and starts to grow like \form{nlsol},
which has a peak at $mt_s \simeq 23 \lsim mt_{\rm sb} \simeq 26$, see
Fig.~\ref{peaktime}. This corresponds to a time slightly {\it earlier}
that the time of symmetry breaking. This is of course natural since,
as we have described, the Higgs field has an inhomogeneous spatial
distribution.  The mean field (coarse-grained over a horizon-sized
volume) is much lower than a typical peak of the field. The top of the
peak follows very approximately the homogeneous equation \form{homo},
with solution \form{nlsol}. High peaks will reach the symmetry
breaking vev much earlier than the mean field, and will oscillate
around the vev with a much larger amplitude that the average
(coarse-grained) field.

Obviously, the phenomenological damping of oscillations that we have
described has to arise from the gradient terms which we have
neglected. Hence, we will improve our approximation by keeping these
terms, but assuming spherical symmetry ($\phi(\x,t) \rightarrow
\phi(r,t)$) around the center of the lump ($r=0$). Our lattice data
support the approximate validity of this assumption. This will allow
us to track the time-evolution of the lump profile as it develops into
bubbles. The two-dimensional partial differential equation for
$\phi(r,t)$ becomes:
\be\label{partialrt}
\ddot\phi(r,t) -  \phi''(r,t) -{2\over r} \phi'(r,t)  -
2\phi^2(r,t) + 2\phi^3(r,t) = 0\,.
\ee 
We have solved this equation numerically. The initial condition was
fixed at a time $mt_{\rm nl}$ when the profile matches expression
\form{peak}.  In order to compare with the non-linear lattice
simulations, we added by hand a tail a long distances , to match the
lattice initial conditions, see Fig.~\ref{lump}.  To fix a unique
solution, one has also to fix $\dot{\phi}(r,t_{\rm nl})$.  Choosing
this derivative equal to zero we have obtained the shape $\phi(r,t)$
for all times in the region of interest.  In Fig.~\ref{bubbles} we
present the result of the comparison of these results with those
obtained from the full 4D lattice real-time equations of motion. The
general shape is quite properly reproduced.  Furthermore, the
oscillations of the peak height are also recovered, see
Fig.~\ref{peaktime}.

As a last remark, note that the bubbles that appear here are {\em not}
vacuum bubbles like those produced in a first order phase transition,
since the interior of them is {\em not} in the true vacuum.
Furthermore, we note also that the ridges of the bubbles are moving
very fast and presumably subsequent collisions between bubbles formed
at different space-time points are highly relativistic, and may be
responsible for a large density of gravitational waves, which could be
seen in LISA.

For a typical lattice configuration 
one can follow the evolution of the Higgs from the formation of the
first bubbles to the breaking of the symmetry with the .gif file
that can be found in the web page:
{\tt http://lattice.ft.uam.es/SymBrk/2dHiggs.gif}

\section{Conclusions}

In this paper we have studied the evolution of a hybrid inflation
model from the quantum false vacuum state at the end of inflation to
the broken symmetry true vacuum state.  A full description of this
dynamics amounts to a non-perturbative, non-linear, real-time
evolution of the quantum system, which looks a priori like a
formidable task.  The size of non-linear effects is given by $\lambda
\phi^2$, where $\lambda$ is the coupling constant and $\phi^2$ the
square of the typical value of the Higgs field. Since initially
$\phi^2$ and $\lambda$ are small, it is reasonable to assume that
perturbation theory is a good approximation and the dynamics is well
approximated by the gaussian Hamiltonian. However, the quantum
evolution of this gaussian system, which can be treated exactly, is
far from trivial. The complexity results from the negative
time-dependent mass-square of low-momentum Higgs modes induced by the
coupling to the inflaton. This tachyonic dynamics generates a faster
than exponential tachyonic growth of low-lying momentum modes of the
Higgs, giving rise to regions where $\lambda \phi^2$ is non-negligible
and where non-linearities set in. In this paper we have shown that the
dynamics of the tachyonic modes is well-described by that of a
classical gaussian random field, a result that holds even after
including perturbative corrections in the coupling, which are still
accessible to exact computation. At this stage important
considerations set in through the appearance of ultraviolet
divergences. High-momentum modes cannot be neglected but their effect
can be absorbed in the value of the couplings of the theory.  Here, in
addition to the usual standard time-independent renormalization, a
renormalization of the initial velocity of the inflaton field is
required to get rid of the time-dependent infinities generated at
first order in the coupling $\lambda$.

The previous analysis justifies the next stage of the study carried
out in this paper, namely the classical non-linear evolution of the
resulting classical field. This problem can be addressed numerically
by formulating the problem on a spatial lattice and evolving the
system according to the classical real-time evolution equations. The
initial conditions on the classical field are determined by the
previously computed (non-self-interacting) quantum Higgs
evolution. Our results are independent of all cut-offs introduced by
this numerical procedure: the initial time of the simulation, the
lattice-spacing and the finite lattice volume.  This, of course,
provided they are taken in the appropriate ranges.

The resulting non-linear evolution which drives the system towards
symmetry breaking is fairly non-trivial. The inhomogeneous Higgs field
distribution has lumps in space whose height grows with time during the
approximately linear evolution phase. This growth continues, although at
a slower pace, when the non-linear terms become relevant. The behaviour
changes again as the highest lumps reach the magnitude of the Higgs
vacuum expectation value. Then the lumps evolve into approximately
spherically symmetric bubbles which expand at a very high speed. It is important not
to confuse these bubbles with those appearing in a first order phase
transition which separate two different phases. Our bubbles are rather like
spherical shock waves as those appearing in Ref.~\cite{GBKLT}. 
These stages of the non-linear evolution can be 
qualitatively and quantitatively understood analytically. The last phase of 
evolution arises as neighbouring bubbles collide and generate higher momentum
modes. This phase is harder to tackle analytically but its early stages,
at least, seems relatively safe for our lattice numerical procedure. 

In the early stages of evolution our results resemble those obtained for a one
component Higgs model in Ref.~\cite{CPR}, as expected from the decoupling of
the different components of the Higgs field in the linear regime.
At later times, however, the comparison is difficult due to the different nature of 
the defects in both theories.

The authors are presently studying how the previously described processes 
might be influenced by the coupling to gauge fields, and its application to the study
of physical phenomena such as baryogenesis. We anticipate that
there is no essential obstruction for incorporating gauge fields, although the
formalism complicates considerably. Furthermore, the numerical evolution
including gauge fields does not change substantially the gross features
of the picture described here. All this will be the subject of a future publication.

\section*{Acknowledgements}

It is a pleasure to thank Julien Lesgourgues for very enlightening
discussions on the quantum to classical transition of fluctuations
during inflation, and for pointing out Ref.~\cite{LPS}. It is also a
pleasure to thank Jos\'e L. F. Barb\'on, Andrei Linde, Gary Felder and
Ester Ruiz Morales for useful comments to the manuscript. This work
was supported in part by the CICYT project FPA2000-980. J.G.B. is on
leave from Universidad Aut\'onoma de Madrid and has support from a
Spanish MEC Fellowship.

\appendix

\section{The formalism of squeezed states}

In this appendix we will summarise the concept of squeezed states so
often used in quantum optics, and recently applied to the study of
quantum fluctuations from inflation \cite{PS,KPS,LPS}.

The canonical harmonic oscillator system \form{Hamiltonian} is
described by two complex functions $(f_k,\ g_k)$, plus a Wronskian
constraint \form{wrk}, and thus we can describe the system in terms of 
three real functions in the standard parametrisation for squeezed states,
\bea\label{ukrw}
u_k(\t) &=& \dk\,[k\,f_k(\t) + g_k(\t)] = 
e^{-i\,\theta_k(\t)}\,\cosh\,r_k(\t)\,,\\ \label{vkrw}
v_k(\t) &=& \dk\,[k\,f_k^*(\t) - g_k^*(\t)] = 
e^{i\,\theta_k(\t) + 2i\,\phi_k(\t)}\,\sinh\,r_k(\t)\,,
\eea
where $r_k$ is the squeezing parameter, $\phi_k$ the squeezing angle,
and $\theta_k$ the phase.

We can also write its relation to the usual Bogoliubov
coefficients, $\{\alpha_k,\ \beta_k\}$
\be
u_k = \alpha_k\,e^{-ik\t}\,, \hspace{2cm}
v_k^* = \beta_k\,e^{ik\t}\,,
\ee
which is useful for the adiabatic expansion, and allows one
to write the average number of particles and other quantities,
\bea\label{nkw}
n_k &=& |\beta_k|^2 = |v_k|^2 = {1\over2k}\,|g_k|^2 + 
{k\over2}\,|f_k|^2 - \half = \sinh^2\,r_k\,,\\
\sigma_k &=& 2\Re\Big(\alpha_k^*\beta_k\,e^{2ik\t}\Big) = 
2\Re\Big(u_k^*\,v_k^*\Big) = \cos2\phi_k\,\sinh2r_k\,,\\
\tau_k &=& 2\Im\Big(\alpha_k^*\beta_k\,e^{2ik\t}\Big) = 
2\Im\Big(u_k^*\,v_k^*\Big) = -\sin2\phi_k\,\sinh2r_k\,.
\eea

We can invert these expressions to give $(r_k,\ \theta_k,\
\phi_k)$ as a function of $u_k$ and $v_k$,
\bea
\sinh r_k &=& \sqrt{{\Re v_k}^2 + {\Im v_k}^2}\,, \hspace{1cm}
\cosh r_k = \sqrt{{\Re u_k}^2 + {\Im u_k}^2}\,, \\
\tan\theta_k &=& -\,{\Im u_k\over\Re u_k}\,, \hspace{1cm}
\tan(\theta_k+2\phi_k) = {\Im v_k\over\Re v_k}\,, \\
\tan2\phi_k &=&  {\Im v_k\,\Re u_k + \Im u_k\,\Re v_k\over
\Re v_k\,\Re u_k - \Im u_k\,\Im v_k}\,.
\eea

Let us now use the squeezing formalism to describe the evolution
of the wave function. The equations of motion for the squeezing
parameters follow from those of the field and momentum modes,
\bea
r_k' &=& {w'\over w}\,\cos2\phi_k\,,\\
\phi_k' &=& - k - {w'\over w}\,\coth2r_k\,\sin2\phi_k\,,\\
\theta_k' &=& k + {w'\over w}\,\tanh2r_k\,\sin2\phi_k\,,
\eea
where we have replaced the time-dependent mass \form{masst} with the 
function $w$, with
\be
w'' = \t\,w \,, \hspace{1cm} \longrightarrow \hspace{1cm}
w(\t) = \bi(\t) + \sqrt3\,\ai(\t) \,,
\ee
with $\ai$ and $\bi$ the two independent Airy functions~\cite{AS},
satisfying $w'(0)=0$. 

\begin{figure}
\includegraphics[width=10cm,angle=-90]{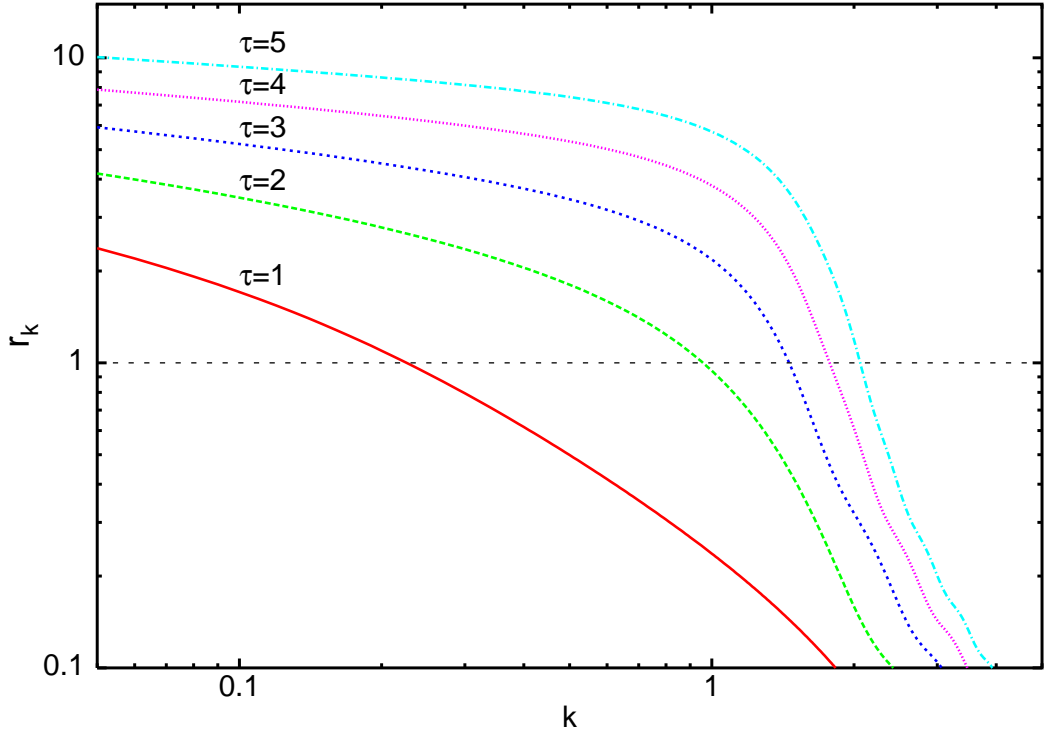}
\caption{The squeezing
parameter at different times in the quantum evolution, as a function of 
wavenumber $k$. Note that at symmetry breaking $\tsb \simeq 5$, the
squeezing parameter is of order $r_k \sim 10$ for long wavelength modes.}
\label{rk}
\end{figure}

As we can see in Fig.~\ref{rk}, the evolution is driven towards large
$r_k \gg 1$. Thus, in that limit, 
$$(\theta_k+\phi_k)' = - {w'\over w}\, {\sin2\phi_k\over\sinh2r_k} 
\to 0\,,$$
and therefore $\theta_k+\phi_k \to$ const. We can always choose this
constant to be zero, so that the real and imaginary components of the
field and momentum modes become
\bea
f_{k1} = \dk\,e^{r_k}\,\cos\phi_k\,, \hspace{1.6cm}
f_{k2} = \dk\,e^{-r_k}\,\sin\phi_k\,, \\
g_{k1} = \kd\,e^{-r_k}\,\cos\phi_k\,, \hspace{2cm}
g_{k2} = \kd\,e^{r_k}\,\sin\phi_k\,.
\eea
It is clear that, in the limit of large squeezing $(r_k\to\infty)$, 
the field mode $f_k$ becomes purely real, while the momentum mode
$g_k$ becomes pure imaginary. This means that the field and 
momentum operators \form{ypkt0} become, in that limit,
\be\left.\ba{l}
\hat y(\k,\t) \ \to \ \sqrt{2k}\,f_{k1}(\t)\,\hat y(\k,\t_0)\\[2mm]
\hat p(\k,\t) \ \to \ \sqrt{2k}\,g_{k2}(\t)\,\hat y(\k,\t_0)\ea
\right\}\ \Rightarrow \hspace{1cm} \hat p(\k,\t) \ \to \ 
{g_{k2}(\t)\over f_{k1}(\t)}\,\hat y(\k,\t)\,.\label{resw}
\ee
As a consequence of this squeezing, information about the initial
momentum $\hat p_0$ distribution is lost, and the positions (or
field amplitudes) at different times commute,
\be
\Big[\hat y(\k,\t_1),\ \hat y(\k,\t_2)\Big] \to \half e^{-2r_k}\,
\cos^2\phi_k \approx 0\,.
\ee
The last result defines what is known as a quantum non-demolition
(QND) variable, which means that one can perform successive
measurements of this variable with arbitrary precision without
modifying the wave function. Note that $y = \delta\phi$ is the
amplitude of fluctuations of the Higgs field after inflation, so what
we have found is: first, that the amplitude is distributed as a
classical gaussian random field with probability \form{probw}; and
second that we can measure its amplitude at any time, and as much as
we like, without modifying the distribution function.

In a sense, this problem is similar to that of a free non-relativistic
quantum particle, described initially by a minimum wave packet, with
initial expectation values $\langle x\rangle_0 = x_0$ and $\langle
p\rangle_0 = p_0$, which becomes broader by its unitary evolution, and
at late times $(t \gg mx_0/p_0)$ this gaussian state becomes an exact
WKB state, $\Psi(x) = \Omega_R^{-1/2}\, \exp(-\Omega\,x^2/2)$, with
$\Im\Omega \gg \Re\Omega$ (i.e. high squeezing limit). In that limit,
$[\hat x,\ \hat p] \approx 0$, and we have lost information about the
initial position $x_0$ (instead of the initial momentum like in our
case), $\hat x(t) \to \hat p(t)\,t/m = p_0t/m$ and $\hat p(t) =
p_0$. Therefore, not only $[\hat p(t_1),\ \hat p(t_2)] = 0$, but also,
at late times, $[\hat x(t_1),\ \hat x(t_2)] \approx 0$.

\section{The Wigner function}

The Wigner function is the best candidate for a probability density of
a quantum mechanical system {\rm in phase-space} \cite{Wigner}.
Of course, we know from Quantum Mechanics that such a probability 
distribution function cannot
exist, but the Wigner function is just a good approximation to that
distribution. Furthermore, in the case of a gaussian state, this
function is positive definite, and can in fact play the role of a
classical probability distribution for the quantum state.

\begin{figure}
\includegraphics[width=10cm,angle=-90]{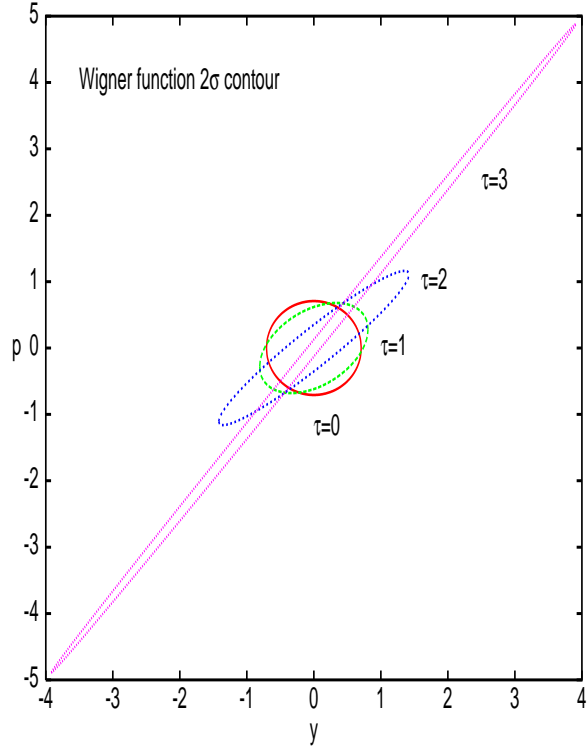}
\caption{
The 2$\sigma$ contour of the Wigner function \form{wignerw} for the mode
  $k=1$, at times $\t=0, 1, 2, 3$. It is clear that, as time
  progresses, the ellipse \form{ellipsew} becomes more elongated
  (squeezed), without changing its area, while the main axis rotates
  counter-clockwise.}
\label{wigner}
\end{figure}

Consider a quantum state described by a density matrix $\rho$.
Then the Wigner function can be written as
\be
W(y_k^0,{y_k^0}^*,p_k^0,{p_k^0}^*) = \int\int\,{dx_1\,dx_2\over
(2\pi)^2}\,e^{-i(p_1\,x_1 + p_2\,x_2)}\,\left\langle y-{x\over2},\t
\right|\rho\left|y+{x\over2},\t\right\rangle\,.
\ee
If we substitute for the state our vacuum initial condition $\rho =
|\Psi_0\rangle\langle\Psi_0|$, with $\Psi_0$ given by the gaussian
wave function \form{wftw}, we can perform the integration explicitly
to obtain
\bea\label{wignerw}
W_0(y_k^0,{y_k^0}^*,p_k^0,{p_k^0}^*) = {1\over\pi^2}\,\exp\left(
-{|y|^2\over|f_k|^2} - 4|f_k|^2\Big|p - {F_k\over|f_k|^2}\,y\Big|^2
\right) \equiv \Phi(y_1,\,p_1)\,\Phi(y_2,\,p_2) \hspace{2mm} \nonumber\\
\Phi(y_1,\,p_1) = {1\over\pi}\,\exp\left\{-\Big({y_1^2\over|f_k|^2}
+ 4|f_k|^2\,\bar p_1^2\Big)\right\}\,, \hspace{1cm}
\bar p_1 \equiv p_1 - {F_k\over|f_k|^2}\,y_1\,. \hspace{1cm}
\eea
However, at time $\t=\t_0$, we have $y_1^0 = \dk = |f_k(\t_0)|\,,
\ p_1^0 = \kd = (2|f_k(\t_0)|)^{-1}$, and $F_k(\t_0) = 0$, so that
$\bar p_1^0 = p_1^0$, and therefore $W_0$ describes a symmetric 
gaussian in phase space, with the same dispersion in both $y$ and 
$p$-directions. The $2\sigma$ contours of this distribution satisfy
\be\label{ellipsew}
{y_1^2\over|f_k|^2} + 4|f_k|^2\,\bar p_1^2 \ \leq \ 1 \hspace{5mm}
\longrightarrow \hspace{5mm}
{y_1^2\over{y_1^0}^2} + {p_1^2\over{p_1^0}^2} \leq 1\,, \hspace{5mm}
{\rm for} \ \t=\t_0 \,,
\ee
which is a circle in phase space. On the other hand, for time 
$\t \gg \t_0$, we have
\bea
|f_k| &\to& \dk\,e^{r_k} \sim y_k^0\,e^{r_k}\,, \hspace{1.5cm}
{\rm growing\ mode} \,, \hspace{1cm} \\
{1\over2|f_k|} &\to& \kd\,e^{-r_k} \sim p_k^0\,e^{-r_k}\,, \hspace{1cm}
{\rm decaying\ mode} \,, \hspace{1cm} 
\eea
so that the ellipse \form{ellipsew} becomes highly ``squeezed'', see
Fig.~\ref{wigner}. Note that Liouville's theorem implies that the
volume of phase space is conserved under Hamiltonian (unitary)
evolution, so that the area within the ellipse should be conserved,
and as a consequence there is no entropy production in this process.
As the probability distribution compresses (squeezes) along the
$p$-direction, it expands along the $y$-direction. At late times, the
Wigner function is highly concentrated around the region
\be
\bar p^2 = \Big(p - {F_k\over|f_k|^2}\,y\Big)^2 < {1\over4|f_k|^2} 
\sim p_0^2\,e^{-2r_k} \ll 1\,.
\ee
We can thus take the above squeezing limit in the Wigner function 
\form{wignerw} and write the exponential term as a Dirac delta function,
\be\label{wignerdelta}
W_0(y,p) \stackrel{r_k\to\infty}{\longrightarrow} {1\over\pi^2}\,
\exp\Big\{\!-{|y|^2\over|f_k|^2}\Big\}\ 
\delta\Big(p - {F_k\over|f_k|^2}\,y\Big)\,.
\ee
In this limit we have
\be
\hat p_k(\t) = {F_k\over|f_k|^2}\,\hat y_k(\t) \ \to \
{g_{k2}(\t)\over f_{k1}(\t)}\,\hat y_k(\t)\,,
\ee
so we recover the previous result \form{resw}. This explains why we 
can treat the system as a classical gaussian random field: the 
amplitude of the field $y$ is uncertain with probability distribution
\form{probw}, but once a measurement of $y$ is performed, we can
automatically assign to it a {\em definite} value of the momentum,
according to \form{resw}.

Note that the condition $F_k^2 \gg 1$ is actually a condition between
operators and their commutators/anticommutators. The Heisenberg
uncertainty principle states that $$\Delta_\Psi A\,\Delta_\Psi B
\geq \half\,\Big|\langle\Psi|[A,\ B]|\Psi\rangle\Big|\,,$$ for any
two hermitian operators (observables) in the Hilbert space of the wave
function $\Psi$. In our case, and in Fourier space, this corresponds to 
\be
\Delta_\Psi y^2(k)\,\Delta_\Psi p^2(k) = F_k^2(\t) + {1\over4}
\geq {1\over4}\,
\Big|\langle\Psi|[y_k(\t),\ p^\dagger_k(\t)]|\Psi\rangle\Big|^2\,,
\ee 
with $|\Psi\rangle = |0,\t_0\rangle$ the vacuum wave function.
On the other hand, $F_k$ can be written as
\be
F_k = \half\,\langle\Psi|\hat p(\k,\t)\,\hat y^\dagger(\k,\t) + 
\hat y(\k,\t)\,\hat p^\dagger(\k,\t)|\Psi\rangle
= -\,{i\over2}\,(g_k\,f_k^* - f_k\,g_k^*) = \Im(f_k^*\,g_k)\,,
\ee
where we have used \form{ypkt0} and $\ a(\k,\t_0)|\Psi\rangle = 0, \ 
\forall \k$. The above relation just indicates that, for any state
$\Psi$, the condition of classicality $(F_k \gg 1)$ is satisfied
whenever, for that state,
\be
\langle\{\hat y_k(\t),\ \hat p_k^\dagger(\t)\}\rangle \gg 
\langle|[\hat y_k(\t),\ \hat p_k^\dagger(\t)]|\rangle = 
\hbar \equiv 1\,.
\ee
It is this condition which allows
one to substitute quantum averages of arbitrary functions $G$ of the
position and momentum operators by classical ensemble averages of the
same function $G$, weighted with the Wigner probability distribution
function, or schematically,
\bea
\langle\Psi|G(\hat y_k,\ \hat p_k)|\Psi\rangle &=&
\int dy_k\,dp_k\,G(y_k, \,p_k)\,W_0(y_k, \,p_k) + 
{\cal O}(\hbar)\\
&\stackrel{r_k\to\infty}{\longrightarrow}& {1\over\pi^2}
\int dy_k\,G\Big(y_k, \,{F_k(\t)\over|f_k|^2}\,y_k\Big)\,
e^{\displaystyle -{y_k^2\over|f_k(\t)|^2} }\,,
\eea
where we have used Eq.~\form{wignerdelta}. As long as $F_k(\t)\gg1$, we can
describe the evolution of our quantum system as that of a classical
gaussian random field. Note that, in this limit, we can ignore the 
normal ordering of the operators in $G(\hat y_k,\,\hat p_k)$.

\section{Non-linear evolution in perturbation theory}

In this appendix we will give details of how to perform perturbative
calculations of the non-linear evolution of our quantum system. We
will also illustrate the perturbative expansion of the correlation
functions of a classical random field. To simplify the expressions
we will consider the case of a single component real scalar(Higgs)
field, but generalisation to complex or multiple component case is
straightforward.

Our goal is to compute the expectation values of products of fields at
different points:
\begin{equation}
\langle\phi(\t_1,\x_1) \ldots \phi(\t_n,\x_n)\rangle
\end{equation}
(In this section we will use the symbol $\phi$ instead of $y$ for the
Higgs field). Here $\phi(\t,\x)$ denotes the Heisenberg picture
field operator, whose relation to the Schr\"odinger picture one
$\phi_s(\x)$ is as follows:
\begin{equation}
\phi(\t,\x)= {\cal U}^\dagger(\t)\, \phi_s(\x)\, {\cal U}(\tau)
\end{equation}
where ${\cal U}(\t)$ is the evolution operator, satisfying:
\begin{equation}
{\cal U}'(\t)=-i{\mathbf H}\, {\cal U} (\t)
\end{equation}
where the prime stands for derivative with respect to $\t$ and
${\mathbf H}$ is the full Hamiltonian. Notice that since the
Hamiltonian depends explicitly on time, the evolution operator cannot
be written as $\exp\{-i \t {\mathbf H}\}$.  If we set $\lambda$ to
zero we get the quadratic Hamiltonian ${\mathbf H}_0$ considered in
the gaussian approximation. The corresponding evolution operator is
${\cal U}_0(\t)$.  Now we go over to the interaction representation by
writing:
\begin{equation}
{\cal U}(\t)= {\cal U}_0(\t)\,  {\mathbf \Omega}(\t)
\end{equation}
where ${\mathbf \Omega}(\t)$ is the characteristic Moller type
operator which satisfies the equation:
\begin{equation}
\label{omegaeq}
 {\mathbf \Omega}'(\t)=-i \,  {\cal U}_0^\dagger(\t)\,   \Hint\,
{\cal U}_0(\t)\,  {\mathbf \Omega} = -i \,   \Hint^{(0)}(\t)\,
{\mathbf \Omega}
\end{equation}
where $\Hint^{(0)}(\tau)$ is the interaction hamiltonian in the
interaction representation. The equation for ${\mathbf \Omega}$ can be
solved in terms of the time-ordered exponential:
\begin{equation}
{\mathbf \Omega}(\tau)=T\exp\{-i\!\int_0^\tau\!dt\,\Hint^{(0)}(t)\}
\end{equation}
This can be used to express the Heisenberg representation fields in terms
of the (gaussian) interaction representation fields:
\begin{equation}
\phi(\t,\x)=T'\exp\{i\!\int_0^\t\!dt\,\Hint^{(0)}(t)\}\;
\phi_0(\t,\x)\ T\exp\{-i\!\int_0^\t\!dt\,\Hint^{(0)}(t)\}
\end{equation}
In the $T$-exponential time grows from right to left and in the
$T'$-exponential left to right.  To obtain the perturbative expansion
one has to expand the $T$-exponential and $\Hint^{(0)}$ in powers of
$\lambda$. The latter has the form
\be
\Hint^{(0)}(\t)= \lambda \int\!d^3x\,\left( \frac{1}{4}\phi_0^4(\t,\x) -
\frac{3 (\delta_1+\delta_2\t)}{2}\phi_0^2(\t,\x) \right) + O(\lambda^2)
\ee
where the second piece is the counter-term needed to renormalise
to this order. Then, substituting the expression of the field inside
the expectation values, everything reduces to expectation values of
products of interaction representation fields $\phi_0(\tau,\x)$.
The latter reduce, by  Wick's theorem,  to products of two-point
functions:
\begin{equation}
\label{freeprop}
G^{(0)}(\tau,\tau',\x-\x')\equiv
\langle\phi_0(\tau,\x)\phi_0(\tau',\x')\rangle=\int
\frac{d^3k}{(2 \pi)^3} e^{i\k(\x-\x')}\,f_k(\tau)f^*_k(\tau')\,.
\end{equation}
This can be decomposed into a real and imaginary part. The real part
corresponds to the expectation value of the symmetrised product, which
in the gaussian theory was chosen to match with the correlation
function of the classical random field. The imaginary part is
proportional to the commutator of the fields, which is a c-number.

We can illustrate the procedure by computing the  2-point function
\be
\langle \phi(\t,\x)  \phi(\t',\x')\rangle= \int
\frac{d^3k}{(2 \pi)^3} e^{i\k(\x-\x')}\,\hat{G}(k,\t,\t')
\ee
to order $\lambda$. Substituting the expression of the Heisenberg field
for the $\t >\t'$ case we get
\begin{eqnarray}\nonumber
\langle T'\exp\{i\!\int_0^\tau\hspace{-1.5mm}dt\,\Hint^{(0)}(t)\}\,
\phi_0(\x,\tau)\
T\exp\{-i\!\int_{\tau'}^\tau\hspace{-1.5mm}dt\,\Hint^{(0)}(t)\}\,
\phi_0(\x',\tau')\, T\exp\{-i\!\int_0^{\tau'}\hspace{-2.2mm}dt\,
\Hint^{(0)}(t)\}\rangle
\end{eqnarray}
In case  $\t' >\t$ the factor sitting between the two fields has to be
replaced by:
\begin{eqnarray}
T'\exp\{i\!\int_\tau^{\tau'}\hspace{-2mm}dt\,\Hint^{(0)}(t)\}
\end{eqnarray}
Notice the peculiar time-ordering of the operators which differs from
the customary perturbative evaluation of (Feynman) Green functions,
which are time-ordered products of field operators.

To do the calculation to order $\lambda$ it is better to start by
expressing the Heisenberg field to this order:
\begin{eqnarray}
\phi(\tau,\x)&=&\phi_0(\tau,\x)+i\int ds\,
[\Hint^{(0)}(s),\phi_0(\tau,\x)] + \ldots \nonumber \\
&=&\phi_0(\tau,\x)+i\lambda \int\!ds\, d^3z\;
[\phi_0(s,\z),\phi_0(\tau,\x)]
\left(\phi_0^3(s,\z) - 3 (\delta_1 + \delta_2 s) \phi_0(s,\z)\right)
\hspace{5mm}
\end{eqnarray}
Finally one obtains
\begin{eqnarray}\nonumber
\hat{G}(k,\t,\t')=f_k(\t)f^*_k(\t')+
6 \lambda \int_0^\t ds\,A_{\mbox{\tiny ren}}(s)\,
\Im[f_k(\t)f_k^* (s)]\ f_k(s)f_k^* (\t')\\
+ 6 \lambda \int_0^{\t'} ds\,A_{\mbox{\tiny ren}}(s)\,
\Im[f_k(\t')f_k^*(s)]\ f_k(\t)f_k^*(s)\,.
\end{eqnarray}
The meaning of $A_{\mbox{\tiny ren}}(\t)$ is given in the main text,
where one can also find the symmetrised Green function.
Repeating the calculation for a Higgs field with $N_c$ real components
one gets the same expression for each component replacing $6$
by $2(N_c+2)$.

We can  compare with the classical evolution. We will use the same symbol
for the classical field $\phi(\t,\x)$. The equations of motion
\begin{equation}
\phi''(\t,\x)= \Delta \phi(\t,\x) +\t\phi(\t,\x) -\lambda
\phi^3(\t,\x)
\end{equation}
can be solved in perturbation theory in $\lambda$. The expansion is
given in terms of tree graphs with lines associated to the retarded
propagator:
\begin{equation}
G_{\mbox{\tiny ret}}(\t,\t',\x-\x')=
-2  \theta(\t-\t')\, \int \frac{d^3k}{(2 \pi)^3} e^{i
\k(\x-\x')}\, \Im[f_k(\t)f^*_k(\t')] \,.
\end{equation}
If we now take $\phi(\t=0,\x)$, and $\pi(\t=0,\x)$, to be
gaussian random fields, then the field at any other time becomes a
non-gaussian random field.  The correlation functions of this field
can be computed in perturbation theory by combining the
aforementioned expansion
 involving the retarded propagator and the expectation
value of gaussian random field.  To match with the quantum calculation
at zero-order in $\lambda$, this has to be taken as the symmetrised
version of Eq.~\ref{freeprop} (this is just given by the Fourier
transform of the symmetric part of $\Sigma$):
\begin{equation}
G^{(0)}_{\mbox{\tiny gauss}}(\t,\t',\x-\x')=
\int \frac{d^3k}{(2 \pi)^3} e^{i\k(\x-\x')}\,\Re[f_k(\t)f_k^*(\t')]\,.
\end{equation}
Notice that terms in the expansion can be associated with Feynman-type
graphs, with modified rules involving two propagators
($G^{(0)}_{\mbox{\tiny gauss}}$ and $G_{\mbox{\tiny ret}}$). These, up
to factors (including Heaviside $\theta$) coincide with the real and
imaginary parts of the quantum propagator.  Indeed, the calculation of
the two-point correlation function to order $\lambda$ matches exactly
with the symmetrised quantum two-point function to this order. For
that one has to apply exactly the same renormalization to the
classical and quantum theories.

Differences can arise to higher order. Essentially, the
retarded-imaginary propagator in the classical theory cannot form
loops by itself. since it arose from the expansion of the field
equations. This need not be the case in the quantum theory. For
example to second order in $\lambda$ there is a contribution to the
two-point function given by the sunset diagram, with three imaginary
propagators joining the two vertices. However, for low-momenta flowing
through the lines (and large enough times) the dramatic difference in
size of the real and imaginary parts of $f_k(\t)f^*_k(\t')$ justifies
that the classical approximation would still be reasonably good.  A
more thorough investigation of these matters is interesting
but exceeds the realm of this paper.

\end{document}